\documentclass[bibauthoryear]{aa}  

\usepackage{graphicx}
\usepackage{txfonts}

\usepackage[font=small]{caption}
\usepackage{float}
\usepackage{soul}

\usepackage{lipsum}
\usepackage{subcaption}         
\usepackage{lscape}             
\usepackage{placeins}           

\usepackage{natbib}
\usepackage{xcolor}
\usepackage[colorlinks=true,citecolor=blue,linkcolor=blue,urlcolor=blue,anchorcolor=blue]{hyperref}

\usepackage[normalem]{ulem}

\begin{document}

\title{Cyclotron lines in subcritical X-ray pulsars: Monte Carlo simulations reveal the origin of the observed variability}


   \author{Prodromos Fotiadis\inst{1,2}
        \and Nick Loudas\inst{3}
        \and Nikolaos D. Kylafis\inst{1,2}
        \and Joachim Tr\"{u}mper\inst{4,5}}

	\institute{
    University of Crete, Department of Physics \& Institute of
		Theoretical \& Computational Physics, 70013 Herakleio, Greece
              \\
             \email{pfotiadis@ia.forth.gr; kylafis@physics.uoc.gr}
		\and
		Institute of Astrophysics,
		Foundation for Research and Technology-Hellas, 71110 Heraklion, Crete, Greece
            \and
            Department of Astrophysical Sciences, Peyton Hall, Princeton University, Princeton, NJ 08544, USA \\
		\email{loudas@princeton.edu}
        \and 
        Max-Planck-Institut f\"{u}r extraterrestrische Physik, 
		Postfach 1312, 85741 Garching, Germany
        \and
        University Observatory, Faculty of Physics, Ludwig-Maximilians Universit\"{a}t, Scheinerstr. 1, 81679 Munich, Germany\\
	}

   \date{Received / Accepted }

 
  \abstract
    {Observed cyclotron resonant scattering features (CRSFs) in X-ray pulsars (XRPs) exhibit strong variability. In the subcritical luminosity regime, the centroid energy ($E_{\rm{CRSF}}$) and line width ($\sigma_{\rm{CRSF}}$) often show positive correlations with the X-ray luminosity.}
    {We investigate the physical origin of the observed variability quantitatively, focusing on the effects of resonant scattering and Doppler shift induced by the plasma flow in the accretion funnel.}
    {We developed a relativistic Monte Carlo code to perform detailed radiative transfer calculations in the accretion funnel above the hotspot and derive angle-dependent spectra. Analytical plasma density and velocity profiles were adopted to account for the effects of radiation pressure on the flow. Approximate resonant scattering cross-sections were employed. We varied the accretion luminosity to explore the resulting variability of the CRSF properties. }
   {
   The emergent spectra exhibit a prominent, asymmetric CRSF accompanied by a broad blue wing. The CRSF is systematically redshifted relative to the classical cyclotron energy, with the magnitude of the redshift decreasing at higher luminosities and for larger viewing angles $\theta$. Both $E_{\rm{CRSF}}$ and $\sigma_{\rm{CRSF}}$ correlate positively with luminosity for all viewing angles. Their absolute values, however, depend strongly on the viewing angle, indicating substantial variability over the pulse cycle and sensitivity to the system geometry. At fixed luminosity, $E_{\rm{CRSF}}$ ($\sigma_{\rm{CRSF}}$) decreases (increases) with increasing $\cos\theta$. Consequently, phase-resolved observations are expected to reveal an anticorrelation between the CRSF centroid energy and width. When applied to the XRP GX 304$-$1, the model reproduces the observed CRSF variability over nearly an order of magnitude in luminosity for geometries in which the accretion funnel is predominantly viewed edge-on.
   }
   {Resonant scattering of radiation propagating through the accreting plasma in the magnetic funnel provides a natural explanation for the observed CRSF variability in subcritical XRPs.}


    \authorrunning{P. Fotiadis et al.}
    \titlerunning{Cyclotron line variability in subcritical X-ray pulsars} 

   \maketitle

\section{Introduction}
Accretion-powered X-ray pulsars (XRPs) are highly
magnetized neutron stars (NSs) in binary systems, accreting from a companion,
typically an OB supergiant or a Be star (for a recent review see \citealt{Weng2024}). Due to the star's
strong magnetic field (usually a few $10^{12}\,\mathrm{G}$), accretion proceeds
along the magnetosphere and the plasma is channeled onto the magnetic poles \citep{Ghosh1979}. This process heats the magnetic polar regions, creates hot spots, and, under certain conditions,
produces accretion columns, where the gravitational energy of the relativistic inflowing plasma is released primarily via hard
X-ray emission \citep{Basko1975, Basko1976, Arons1987}.

A substantial fraction of these sources shows absorption-line-like features in the
$10$--$100\,\mathrm{keV}$ energy range of their spectra; the so-called
cyclotron lines or cyclotron resonant scattering features (CRSFs) \citep[for a review, see][]{Staubert2019}.
CRSFs originate from a quantum-mechanical process called magnetic scattering. The perpendicular momentum of charged particles in a strong magnetic field is quantized, resulting in discrete energy states, known as Landau levels \citep{Landau1930}. Resonant scattering of photons with inflowing electrons in the vicinity of the NS surface imprints the fundamental CRSF in the continuum \citep{Truemper1977, Truemper1978}, with several XRPs also featuring harmonics \citep{Truemper1978, Santangelo1999, Fuerst2018, Yang2023}.
The energy difference between two successive Landau levels (hereafter referred to as the cyclotron energy) in the non-relativistic case is \begin{equation}
E_{\mathrm{c}} \equiv \dfrac{\hbar eB}{m_{\mathrm{e}} c}\approx 11.6 \left( \dfrac{B}{10^{12}\,\mathrm{G}}\right) ~\mathrm{keV}, \label{E_cyc}
\end{equation}
where $B$ is the magnetic field strength in the line forming region, $\hbar$ is the reduced Planck constant, $m_{\mathrm{e}}$ is the electron's rest mass, $c$ is the speed of light, and $e$ is the elementary charge. 
As $E_{\mathrm{c}}$ is linearly proportional to the magnetic field, the detection of CRSFs provides a direct and robust means of measuring the magnetic field strength in XRPs and, more broadly, serves as a diagnostic of the physical conditions near the neutron star surface \cite{Gnedin1974}.

Detailed observations of CRSFs in XRPs have revealed significant variability both with pulse phase and X-ray luminosity $L$ \citep[for reviews, see][]{Staubert2019, Mushtukov2022}. Pulse-phase modulations indicate anisotropic emergent spectra (e.g., \citealt{Maniadakis2025}), whereas variations with $L$ (accretion rate) hint at changes in the emission region \citep[e.g.,][]{Shui2024}. CRSFs also show evidence for long-term variability, as observed in Her X-1 \citep{Staubert&Klochkov2017,Xiao:2019obt}, but the underlying physical mechanism remains unclear.

Of particular interest is the dependence of the observed CRSF centroid energy, $E_{\mathrm{CRSF}}$, on the X-ray luminosity, $L$. Such behavior has been reported in several XRPs, including V0332$+$53 \citep{Tsygankov2006,Tsygankov2010}, GX 304$-$1 \citep{Klochkov2012,Rothschild2017}, and GRO J1008$-$57 \citep{Chen2021}. High-luminosity sources $(L \gtrsim 10^{37}\,\mathrm{erg}\,\mathrm{s}^{-1})$, typically present a negative correlation between $E_{\mathrm{CRSF}}$ and $L$, whereas in low-luminosity sources $(L \lesssim 10^{37}\,\mathrm{erg}\,\mathrm{s}^{-1})$ the trend reverses, with a positive correlation emerging. A complete transition from positive to negative correlation has recently been identified in pulse-to-pulse analyses of 1A 0535+262 by \cite{Shui2024}, further reinforcing this distinction among sources.  The leading physical explanation is the emergence of an accretion column once the luminosity exceeds a critical value $L^* \sim 10^{37} \,\mathrm{erg}\,\mathrm{s}^{-1}$ \citep{Basko1976,Mushtukov2015a}, signaling a transition in the accretion regime \citep{Becker2012}; from a surface-dominated hot spot regime $(L \lesssim L^*)$ to an accretion-column regime $(L \gtrsim L^*)$.

For supercritical (accretion-column-dominated) sources, it has been shown analytically \citep{Basko1976,Becker2007} and through first-principles radiation-magnetohydrodynamic (RMHD) simulations \citep{Klein1996, Zhang2022, Zhang2023, Zhang2025} that the braking of the infalling plasma is caused by an extended radiation-mediated shock (RS) forming in the magnetically confined, optically thick, accretion column. Both analytical studies \citep{Basko1976} and Monte Carlo (MC) approaches \citep{Loudas2023} support the RS as the likely formation site of CRSFs in high luminosity XRPs. This scenario successfully predicts the observed anti-correlation between $E_{\mathrm{CRSF}}$ and $L$: as the accretion rate increases, the accretion column grows in height, while the magnetic field strength drops with altitude. Thus, the cyclotron energy in the line-forming region shifts to lower values at higher luminosities \citep{Burnard1991,Loudas2024b}. An alternative scenario, proposed by \cite{Poutanen2013}, attributes CRSF formation to surface reflection of hard X-rays emitted from the accretion column, but detailed radiative-transfer simulations challenge the viability of this interpretation \citep{Kylafis2021}.

At low luminosities (hot-spot-dominated emission), the outgoing radiation is dynamically unimportant, the accreting plasma reaches the NS surface in free fall, and the formation of CRSFs is believed to occur within the NS atmosphere. Even though no settled theory exists explaining the deceleration of the falling material, two leading mechanisms are found in the literature: 1) braking of plasma in the atmosphere by Coulomb collisions \citep{Zeldovich1969, Miller1987, Sokolova-Lapa}, and 2) emergence of a collisionless shock (CS) where the infalling plasma is thermalized upon passing through the shock \citep{Bisnovatyi1970, Langer&Rappaport1982, Bykov2004}. 
The former model considers spectral formation in the NS atmosphere and can explain the observed double-hump in the hard X-ray spectra of low-luminosity XRPs \citep{Sokolova-Lapa}, but it predicts zero variability of the cyclotron line centroid with luminosity and thus cannot reproduce the observed trends of cyclotron lines. 

Conversely, the CS scenario has been successful in explaining the positive correlation between $E_{\mathrm{CRSF}}$ and $L$ in several sources \citep{Staubert2007,Vybornov2017, Rothschild2017, Roy2025}. In that case, spectral formation takes place between the NS surface and the shock discontinuity. Since the magnetic field strength (and effectively $E_{\rm CRSF}$) increases as the altitude decreases, and the CS height decreases with accretion rate (luminosity $L$) \citep{Bykov2004}, a positive correlation between $E_{\rm CRSF}$ and $L$ arises (see also \citealt{Becker2012}). However, first-principles kinetic simulations are required to address whether the formation of such a shock is feasible under conditions encountered above the NS atmosphere and determine its properties. Once the shock dynamics is understood, subsequent radiative transfer calculations need to be conducted, followed by comparison with observed spectra and pulse profiles.

\cite{Mushtukov2015b} pointed out that for subcritical luminosities ($L \lesssim L^*$) even though the propagation of the hot spot's radiation through the mildly relativistic infalling plasma cannot halt accretion, yet it can significantly modify the velocity and density profiles above the NS surface. This effect was verified by robust RMHD simulations by \cite{Markozov2023}. In this scenario, the formation of the CRSF takes place above the hot spot. Upward-propagating X-rays undergo resonant scattering with the infalling electrons, producing a red-shifted CRSF relative to the local cyclotron energy, due to the Doppler effect between the laboratory frame (hot spot) and the moving electron's bulk frame. This model predicts a positive correlation between $E_{\mathrm{CRSF}}$ and $L$ without postulating the presence of a CS. A qualitative explanation follows. Radiation pressure acts to decelerate the accretion flow, influencing the degree of redshift in the CRSF. As luminosity (and hence radiation pressure) increases, the accretion-flow velocity right above the hot spot decreases, reducing the Doppler shift and shifting the CRSF centroid to higher energies (closer to the local $E_{\mathrm{c}}$). 
Although the model shows promise and has been successful in qualitatively explaining observations of the X-ray pulsar GX 304$-$1, no quantitative calculations of cyclotron line formation and variability have yet been carried out within this framework. 

In this paper, we investigate the aforementioned model by conducting robust MC radiative transfer simulations to obtain the angle-resolved spectrum for a typical X-ray pulsar, focusing on the properties of the emergent CRSF, and studying its dependence on accretion rate. We aim to quantitatively test the following predictions made by \cite{Mushtukov2015b}:

\begin{enumerate}
    \item[(i)] $E_{\rm CRSF}$ is positively correlated with the luminosity $L$.
    
    \item[(ii)] The line width, $\sigma_{\rm CRSF}$, is positively correlated with $L$.
    
    \item[(iii)] Both $E_{\rm CRSF}$ and $\sigma_{\rm CRSF}$ vary significantly over the pulse phase.
\end{enumerate}

The structure of this paper is organized as follows. 
In Sect.~\ref{sec:model}, we introduce the physical model, detailing the 
velocity and density profiles, along with the structure of the accretion channel. In Sect.~\ref{sec:code}, we provide a detailed description of our MC code. 
In Sect.~\ref{sec:results}, we present the simulated spectra for a typical 
X-ray pulsar and analyze their primary features. In Sect.~\ref{sec:comparison}, we compare our MC results with 
observational data from GX~304$-$1 and discuss the degeneracies inherent 
in the methodology. In Sect.~\ref{sec:discussion}, we comment
on the results and draw our conclusions.

\section{The model} \label{sec:model}

In this section, we outline the underlying features of the model developed by \cite{Mushtukov2015b}, which applies to sources in the subcritical luminosity regime ($L \lesssim L^*$).
\subsection{Structure of the accretion funnel}
  
We consider magnetospheric accretion onto the magnetic pole of a highly magnetized NS. The strong magnetic field confines the accretion flow into a funnel, and the plasma is moving along the field lines only. The impact of infalling plasma heats the NS atmosphere at the base of the funnel, forming a hot spot that produces an outgoing photon beam. For simplicity, we adopt a cylindrical geometry for the filled accretion column, although more realistic accretion flow models favor bow-shaped or annular hot spots \citep{Basko1976, Romanova2004, Zhu2025}. Seed photons are emitted isotropically upwards from the base of the cylinder with a blackbody spectrum. 
We further assume a uniform magnetic field $\vec B$ aligned with the magnetic axis, i.e., \(\vec B = B\hat{z} \). This approximation is justified because the vertical extent of the line-forming region and the size of the funnel are, by construction, much smaller than the NS radius.

The interaction between upward-moving photons and the infalling electrons through Compton scattering exerts a drag (radiation-pressure) force on the falling plasma, altering its velocity and density profiles in a non-trivial manner. In the subcritical luminosity regime considered here, the plasma reaches the surface with a non-zero terminal velocity before being completely thermalized via Coulomb collisions in the atmosphere \citep{Zeldovich1969}. Conservation of momentum in the non-relativistic regime for a stationary flow dictates

\begin{equation}
    (\vec u \cdot \nabla) \vec u = \dfrac{1}{\rho}\vec G_{\mathrm{rad}},
    \label{momentum_eqn}
\end{equation}
where $\rho$ and $\vec{u}$ are the plasma density and bulk velocity, respectively, while $\vec G_{\mathrm{rad}}$ is the radiation-pressure force, proportional to the radiation flux times the opacity. In this equation, we neglect the gas pressure and gravity terms, as the radiation-pressure force is the dominant one in the vicinity of the NS surface. \cite{Mushtukov2015b} solved Eq.~\eqref{momentum_eqn} and derived the following analytical expression for the velocity profile $\beta(h)\equiv |u(h)|/c$ in the center of the column, considering isotropic emission from the hot spot, accounting for the dependence of the outgoing radiation flux on the height $h$ above the NS surface, and considering motion along the field lines only,
\begin{equation}
    \beta(h) = \left[\beta_{\rm ff}^2 - \left(\beta_{\rm ff}^2 - \beta_0^2\right)
    \left( 1 - \frac{2}{\pi}\tan^{-1}\left(\frac{h}{d}\right)\right)\right]^{1/2},
    \label{eq:velocity_profile}
\end{equation} 
where $d$ is the radius of the hot spot, estimated by the following power-law scaling \citep{Mushtukov2015b}
\begin{equation}
    d \approx 6\times 10^3 \, \Lambda^{-3/8} \,L_{37}^{9/35}\, B_{12}^{-3/14} \,m^{-81/140} \,R_6^{39/35} \, \mathrm{cm},
    \label{eq:hot spot_size}
\end{equation}
where $\Lambda = 0.5$ is a geometrical parameter related to the coupling of the magnetosphere to the disk \citep{Ghosh1978}, $L_{37}\equiv L / 10^{37}\,\mathrm{erg\,s^{-1}}$, 
$B_{12} \equiv B / 10^{12}\,\mathrm{G}$, $m = M/M_\odot$, and 
$R_6 \equiv R / 10^6\,\mathrm{cm}$. In Eq.~\eqref{eq:velocity_profile}, $\beta_{\mathrm{ff}}$ denotes the free-fall speed, divided by $c$,
of the accreting plasma at the neutron-star surface, while $\beta_0$ represents the terminal velocity at the 
stellar surface ($h=0$). The latter is related to the accretion luminosity through energy conservation, assuming that the kinetic energy of the accreting plasma is instantaneously converted into radiation via thermalization at the hot spot and Compton scattering above it. This yields a scaling relation of the form
\begin{equation}
    \beta_0 = \beta_{\mathrm{ff}}\sqrt{1-\dfrac{L}{L^*}},
    \label{eq:beta_0}
\end{equation}
with the critical luminosity $L^*$ defined as the limiting luminosity at which the infalling material is fully decelerated ($\beta_0=0$) by radiation pressure at the stellar surface; above this threshold, an RS is expected to rise above the hot spot \citep{Basko1976}.

To ensure that the model is self-consistent, the velocity profile described 
by Eq.~\eqref{eq:velocity_profile} necessitates a non-uniform density profile that obeys the conservation of mass
\begin{equation}
    \nabla \cdot (\rho \vec u) = 0.
\end{equation}
Given the cylindrical geometry and a constant accretion rate ($\dot{M} = \mathrm{const}$), the electron number density profile reads \citep{Mushtukov2015b}
\begin{equation}
    n_{\mathrm{e}}(h) \approx 2\times 10^{19} \, L_{37}^{3/5} \,B_{12}^{1/2}\, \beta^{-1}(h)\, \mathrm{cm}^{-3}.
    \label{eq:density_profile}
\end{equation}
Finally, following \cite{Mushtukov2022} (see Sect. 4.3 therein), we estimate the effective temperature of the hot spot using the Stefan-Boltzmann law
\begin{equation}
    k_{\mathrm{B}}T = k_{\mathrm{B}}\left(\dfrac{L}{2 \sigma_{\mathrm{SB}} S_D}\right)^{1/4}\approx 6.3L_{37}^{3/20} 
    \Lambda^{7/32}\,
    m^{13/80}\,
    R_6^{-19/40}\,
    B_{12}^{1/8}\,\mathrm{keV}
    \label{eq:temperature}
\end{equation}
where $\sigma_{\mathrm{SB}}$ is the Stefan-Boltzmann constant and $S_D$ denotes the hot spot area, given by \cite{Mushtukov2015b}
\begin{equation}
    S_D \approx 3\times 10^9\,\Lambda^{-7/8}\,m^{-13/20}\,R_6^{19/10}\,B_{12}^{-1/2}\,L_{37}^{2/5}~\mathrm{cm}^2. 
    \label{eq:hot spot_area}
\end{equation}
The factor of two in the denominator of Eq.~\eqref{eq:temperature} accounts for the existence of two antipodal hot spots.

The approximate power-law scaling adopted for the hot spot effective temperature captures the expected positive correlation with luminosity. However, the absolute normalization of the temperature remains uncertain at the level of a factor of $\sim 2$, depending on the accretion geometry (e.g., the shape of the landing region, stochastic variability, and system inclination) and on the microphysics of the thermalization process in the NS atmosphere (e.g., CS versus Coulomb collisions). Deriving a robust prescription for temperature would require detailed RMHD simulations (such as those of \citealt{Markozov2023}) and is therefore beyond the scope of this work. Nevertheless, we emphasize that our conclusions regarding the CRSF properties and their dependence on luminosity and/or viewing angle are insensitive to the assumed injected photon spectrum, and hence to the specific choice of the hot spot (blackbody) temperature.

\subsection{CRSF formation} 
\label{sec:CRSF_formation}

As the accreting plasma descends towards the surface, upward-propagating photons undergo Compton scattering with the infalling electrons. For typical values of the electron density and hot spot size, the perpendicular Thomson optical depth of the accretion funnel lies in the optically thin regime, 
\begin{equation}
\tau \sim n_{\mathrm{e}} \sigma_{\mathrm{T}} d \lesssim 0.2,
\end{equation} such that most photons escape without scattering. However, magnetic scattering includes resonant processes in which photons with energies near the cyclotron resonance experience a strongly enhanced cross-section (up to $\sim 10^{6} ~\sigma_{\mathrm{T}}$, \citealt{Canuto1971, Ventura1979, Harding1991, Sina1996, Mushtukov2016, Loudas2021}) and are effectively trapped within the medium. These resonant photons dominate the radiative deceleration of the infalling plasma \citep{Mushtukov2015b}: they undergo multiple resonant scatterings until they exchange sufficient energy with the electrons to shift out of resonance and escape the funnel. This process naturally produces an absorption-line-like feature accompanied by two broad, emission-like wings, reflecting photon redistribution in energy rather than true photon absorption, since the total number of photons is conserved in scattering.

In this study, we assume that the magnetic field strength is well below the QED critical value
\begin{equation}
B_{\rm cr} = \dfrac{m_{\mathrm{e}}^2 c^3}{e\hbar} \approx 4.413 \times 10^{13}~\mathrm{G},
\end{equation}
such that the dimensionless field strength $b \equiv B/B_{\rm cr} \ll 1$, and we also consider the infalling electron's temperature $T_{\mathrm{e}}$ to be substantially smaller than the local cyclotron energy, that is, $k_{\mathrm{B}}T_{\mathrm{e}} \ll E_{\mathrm{c}}$. In short, we restrict our analysis to systems that satisfy 
\begin{equation}
    k_{\mathrm{B}}T_{\mathrm{e}} \ll  E_{\mathrm{c}} \ll m_{\mathrm{e}} c^2,
\end{equation}
which is suitable for most XRPs in the hot spot regime \citep{Becker2005}. In this regime, the vast majority of electrons occupy the fundamental Landau level ($n=0$), and thus we can safely account for transitions between the ground state and the first excited Landau level ($0 \rightarrow 1 \rightarrow 0$)  \citep{Nobili2008, Loudas2021}. 

In the infalling electron's rest frame, the relativistic cyclotron resonance energy is given by
\begin{equation}
E_{\mathrm{res}} = m_{\mathrm{e}} c^2 \frac{2b}{1+\sqrt{1+2b\sin^2\theta_{\mathrm{rf}}}}  = \frac{2E_{\mathrm{c}}}{1+\sqrt{1+2b\sin^2\theta_{\mathrm{rf}}}},
\label{res_energy}
\end{equation}
where $\theta_{\mathrm{rf}}$ is the angle between the photon momentum and the magnetic field in the electron frame. For small values of $b$, one would expect the CRSF energy centroid to appear at $E_{\mathrm{res}} \approx E_{\mathrm{c}}$. Due to the bulk motion of the infalling plasma, this resonance energy appears Doppler shifted when viewed from the laboratory frame; thus, it is not photons with energy $\approx E_{\mathrm{res}}$ (in the lab frame) that are in resonance, but those that the Doppler boost to the electron frame brings them in resonance, that is
\begin{equation}
    E \approx \dfrac{E_{\mathrm{res}}}{\gamma(1 + \beta \cos\theta)} < E_{\mathrm{res}},
    \label{redshift_energy}
\end{equation}
where $\theta\in (0,\pi/2)$ is the angle between the photon momentum and the magnetic field in the lab, and $\gamma = (1-\beta^{2})^{-1/2}$. Therefore, the CRSF will not form at $E_{\mathrm{res}}$, but will appear redshifted. As Eq.~\eqref{redshift_energy} indicates, the amount of redshift depends on the plasma's velocity in the region where the interaction takes place, as well as on the photon's direction of propagation. 

Higher luminosities correspond to lower infall velocities (see Eq.~\ref{eq:beta_0}), leading to higher CRSF centroid energies. This results in a positive correlation between the luminosity and the CRSF centroid energy \citep{Mushtukov2015b}. In addition, for a given velocity, the Doppler redshift decreases with increasing viewing (propagation) angle $\theta$, implying that the CRSF centroid energy increases with $\theta$. However, this qualitative picture is based on the single-scattering approximation. Multiple scatterings could induce modulations in the shape and location of the emergent CRSF. Obtaining accurate predictions of the model requires full radiative-transfer calculations.

\section{Monte Carlo code} \label{sec:code}

The relativistic MC radiative transfer code employed in this work is based on the implementation developed by \citet{Loudas2023}, to which the reader is referred for a detailed description. Here, we summarize the main parts of the code and describe the improvements with respect to the version of \cite{Loudas2023}. Radiative transfer is treated using the forced first collision technique, which ensures efficient sampling of photon–electron interactions. Photon packets are injected in the accretion funnel and are followed by the MC code until they escape. A key advantage of the present implementation is its flexibility, as it supports arbitrary density and velocity profiles within the accretion channel, enabling direct tests of different physical scenarios. Moreover, it incorporates an adaptive step size algorithm that increases efficiency, while maintaining accuracy in regions of steep gradients. Figure \ref{fig1} shows an illustration of the system's geometry and the numerical propagation of a photon packet. 

\subsection{Accretion channel and plasma}

The accretion channel is modeled as an axisymmetric cylindrical column of radius $d$ and semi-infinite length, with its base located at the NS surface. The magnetic field is considered to be homogeneous along the $z$ axis, that is, $\vec B=B\hat{z}$, because the height of the line-forming region is much smaller than the NS radius. 
The electron density, $n_{\mathrm{e}}(\vec{r})$, evaluated at $\vec{r}=(x, y, z=h)$ is given by Eq.~\eqref{eq:density_profile}, while the bulk velocity of the infalling plasma is taken to be $\vec{u}(\vec{r}) = -\beta(\vec{r}) c~\hat{z}$, with $\beta(\vec r)$ computed by Eq.~\eqref{eq:velocity_profile}. 
Even though both density and velocity profiles are expected to differ across the column \citep{Markozov2023}, for the purposes of this work, we utilized the analytical 1D profiles derived by \cite{Mushtukov2015b}. Infalling electrons were treated as a cold plasma, an assumption justified by observations \citep{Becker2005}; 
thus we set $T_{\mathrm{e}} =0$ in our simulations. 

\subsection{Seed photons}
\label{sec:seed_photons}

Photon packets are emitted isotropically upward, with their initial positions sampled homogeneously from the base of the accretion channel. In our coordinate system, this base corresponds to the region $
    \mathcal{R} = \{ (x, y, z=h) \in \mathbb{R}^3 : x^2 + y^2 \leq d^2, h = 0 \} $.
Each photon packet is initialized with a specified weight, $w = N$ (typically $10^6$), and is assigned an initial energy $E$ drawn from a blackbody distribution at temperature $T$ (given by Eq.~\ref{eq:temperature})
\begin{equation}
    n(E) ~dE\propto \dfrac{E^2}{e^{E/k_{\mathrm{B}}T}-1}~dE.
    \label{eq:blackbody}
\end{equation}
\subsection{Optical depth calculation and photon propagation}

\begin{figure}[t]
\centering
\includegraphics[width=\hsize]{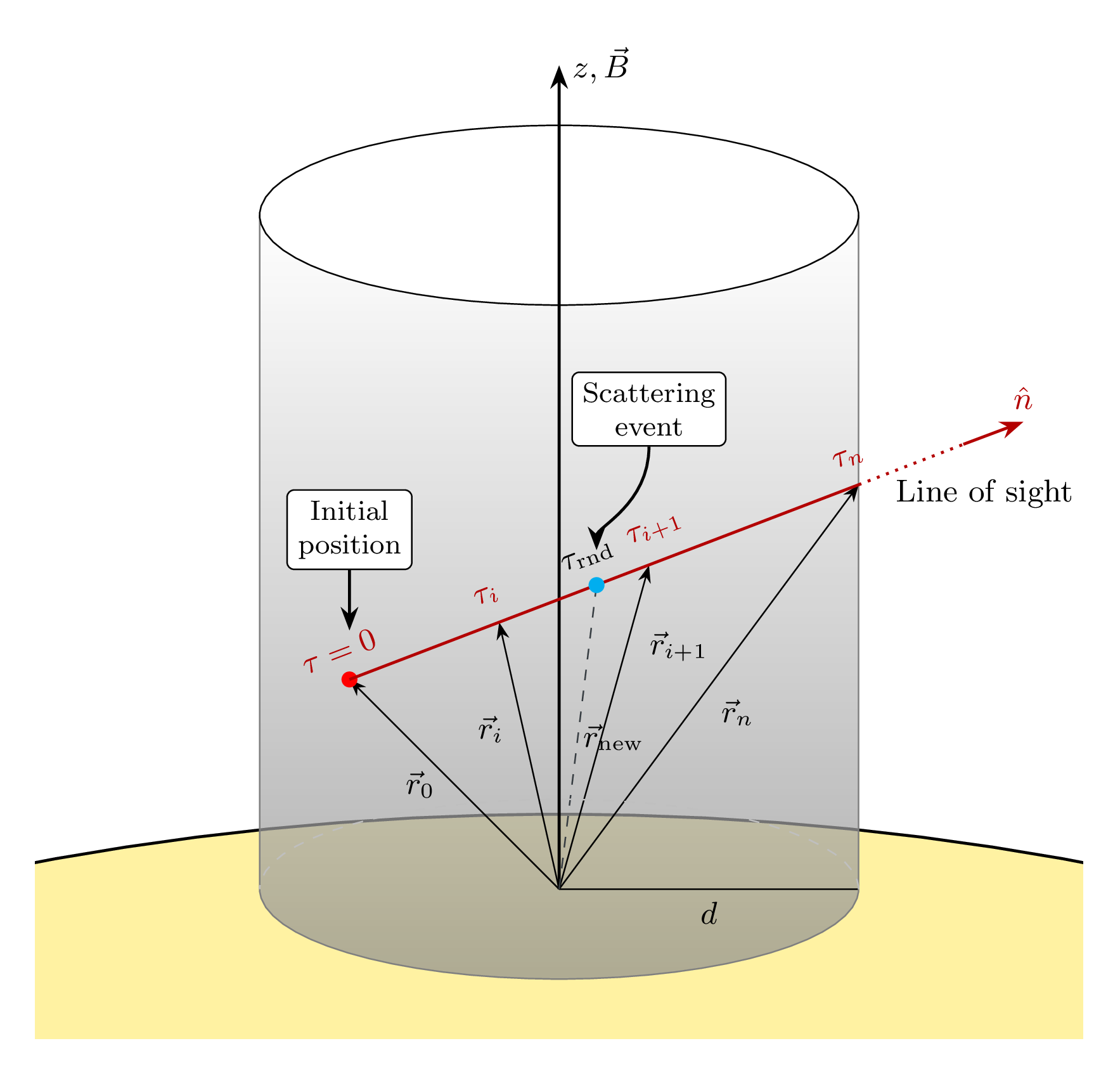}
  \caption{Illustration of the MC code's geometry, coordinates, and calculation of the line-of-sight optical depth. A photon packet located at $\vec r_0$ is moving in direction $\hat n$. The accumulated optical depth is computed by dividing the line of sight into a sequence of points $\vec r_i$ using an adaptive step-size algorithm. A target optical depth $\tau_{\rm rnd}$ is then sampled, and the position of the next scattering event is determined via linear interpolation between the discretized points. We assume a uniform magnetic field because the height of the line-forming region is much smaller than the NS radius. 
  }
     \label{fig1}
\end{figure}

Photon packet propagation is carried out as follows. Consider a photon packet with weight $w$, located at position $\vec{r_0}$ and 
propagating in direction $\hat{n}=(\cos\varphi \sin\theta, \sin\varphi \sin\theta,\cos\theta)$ with energy $E$. Here, $\theta$ is the angle relative to the magnetic axis, while $\varphi$ is the azimuthal one. The cumulative optical depth along the line-of-sight is defined as
\begin{equation}
    \tau_{n}(\vec{r_0}, \hat{n}, E) = \int_{\mathcal C} n_{\mathrm{e}}(\vec{r}) \, 
    \sigma_{\mathrm{eff}}\!\left(E_{\rm rf}, \vec{r}, \hat{n}_{\rm rf}\right) \, ds, 
    \label{eq:tau_along_C}
\end{equation}
where the trajectory $\mathcal C$ is a straight line extending from $\vec{r_0}$ to the cylindrical funnel's boundary and $ds$ is the infinitesimal scalar path length along $\mathcal C$. 
The effective cross-section $\sigma_{\mathrm{eff}}$ accounts for the bulk motion of the plasma and is defined as
\begin{equation}
    \sigma_{\mathrm{eff}}\!\left(E_{\rm rf}, \vec{r}, \hat{n}_{\rm rf}\right) = 
    (1 + \beta(\vec{r})\cos \theta) \, \sigma(E_{\rm rf}, \theta_{\rm rf}),
    \label{eq:sigma_eff}
\end{equation}
where $\sigma$ is the magnetic scattering cross-section expressed in terms of quantities evaluated at the electron rest frame (see Sect. \ref{sec:resonant_scattering}) and $\beta \geq0$, by construction.
The subscript ``rf'' denotes quantities measured in the electron rest frame. The Lorentz transformations relate the rest-frame 
quantities to the laboratory-frame ones as follows
\begin{equation}
    \cos\theta_{\rm rf} = \frac{\beta + \cos\theta}{1 + \beta\cos\theta}, \quad \varphi_{\mathrm{rf}} = \varphi,
    \label{eq:angle_lorentz}
\end{equation}
and
\begin{equation}
    E_{\rm rf} = \gamma E (1 + \beta \cos\theta),
    \label{eq:energy_lorentz}
\end{equation}
where $\gamma=(1-\beta^2)^{-1/2}$.

To determine the location of the next scattering event, $\tau_n$ is computed numerically using an adaptive step-size algorithm to 
handle regions where the electron density $n_{\mathrm{e}}(\vec{r})$ and/or $\beta(\vec{r})$ vary rapidly. The 
integration path is discretized into a sequence of points $\vec{r}_0, \vec{r}_1, \dots, \vec{r}_n$ along $\mathcal C$, 
where $\vec{r}_n$ is the point of the line-of-sight's intersection with the column boundary. Given a point $\vec{r}_i$, the subsequent point $\vec{r}_{i+1}$ is determined as follows. 
First, we calculate the local mean free path
\begin{equation}
    \lambda_{i} \equiv \bar{\lambda} (\vec{r}_i, \hat{n}, E) 
    = \left[ n_{\mathrm{e}}(\vec{r}_i) \, \sigma_{\mathrm{eff}}(E_{\rm rf},\vec{r}_i, \hat{n}_{\rm rf}) \right]^{-1}.
\end{equation}
We then propose a tentative step size equal to 1\% of the local mean free path
\begin{equation}
    |\vec{r}_{i+1} - \vec{r}_i| = 0.01 \, \lambda_i.
\end{equation}
To ensure linearity over this step, we evaluate $\lambda_{i+1}$ at the new location 
and enforce the relative error condition
\begin{equation}
    \frac{|\lambda_{i+1} - \lambda_i|}{\lambda_i} < 0.01. \label{eq:step_test}
\end{equation}
If condition~\eqref{eq:step_test} is violated, the step is rejected and $\vec{r}_{i+1}$ 
is redefined closer to $\vec{r}_i$ by halving the step size (backtracking), that is by increasing the spatial resolution locally, until the 
condition is satisfied
\begin{equation}
    \vec{r}_{i+1} \to \vec{r}_i + \frac{1}{2}(\vec{r}_{i+1} - \vec{r}_i).
\end{equation}
Once the step is accepted, the accumulated optical depth is updated as:
\begin{equation}
    \tau_{i+1} = \tau_{i} + \frac{|\vec{r}_{i+1} - \vec{r}_i|}{\lambda_i},
\end{equation}
with the initial condition $\tau_0 = 0$. This process continues until the boundary 
is reached (or $\tau > 20$ for efficiency), yielding the total optical depth $\tau_n$. A fraction $w e^{-\tau_n}$ of the photon packet escapes the domain, and its energy and direction are recorded.

The remaining undergoes scattering with an electron in the column. The probability to scatter after covering a distance equivalent to optical depth $\tau$ is given by a truncated exponential distribution in $\tau$, thus we sample a target optical depth $\tau_{\rm rnd}$ using
\begin{equation}
    \tau_{\rm rnd} = -\ln\left[ 1 - \xi \left(1 - e^{-\tau_n}\right) \right],
\end{equation}
where $\xi \in [0,1)$ is a uniform random number; $\xi = 0$ implies scattering without moving forward at all, while $\xi \to 1$ indicates that the photon packet will scatter right before it reaches the boundary.  We then locate the index $i$ such 
that $\tau_i \leq \tau_{\rm rnd} < \tau_{i+1}$. The precise location of the next 
scattering event, $\vec{r}_{\rm new}$, is found via linear interpolation along the segment
\begin{equation}
    \vec{r}_{\rm new} = \frac{\vec{r}_i(\tau_{i+1} - \tau_{\rm rnd}) + \vec{r}_{i+1}(\tau_{\rm rnd} - \tau_i)}{\tau_{i+1} - \tau_i},
\end{equation}
where $|\vec{r}_i - \vec r_0| \leq |\vec r _{\rm new} - \vec r_0| < |\vec r _{i+1} - \vec r_0|$.

Once the exact location of the scattering event is specified, we update the weight of the remaining photon packet as follows
\begin{equation}
    w_{\rm new} = w \left(1 - e^{-\tau_n}\right).
\end{equation}
This formula ensures conservation of the total number of photons. If $w_{\rm new}$ is larger than $1$, we carry out the scattering event as described below; otherwise, we proceed to the injection and propagation of a new photon packet (Sect. \ref{sec:seed_photons}).

The scattering process is treated in three steps. First, we Lorentz-transform the photon-packet energy and propagation direction from the laboratory frame to the electron rest frame using the local bulk velocity evaluated at the scattering location $\vec r _{\rm new}$ (Eqs.~\ref{eq:angle_lorentz} and \ref{eq:energy_lorentz}). Next, we sample a new photon propagation direction $\hat n^\prime_{\mathrm{rf}}$ in the electron rest frame employing the differential scattering cross section (see Sect. \ref{sec:resonant_scattering}), and compute the post-scattering photon energy $E^\prime_{\mathrm{rf}}$ via energy conservation (Eq.~(8) in \citealt{Loudas2021}). Finally, we apply the inverse Lorentz transformation to return to the laboratory frame and update the photon-packet properties,
\begin{equation}
    E, ~\hat n,~ \vec r_0, ~w \longleftarrow E^\prime, ~ \hat n^\prime, ~ \vec r_{\rm new}, ~w _{\rm new},
\end{equation}
after which the photon is propagated to the next interaction.

\subsection{Prescription of resonant Compton scattering}
\label{sec:resonant_scattering}

In this work, we focus on the formation of the CRSF, so we need to adopt resonant magnetic scattering cross-sections.
Fully relativistic quantum-electrodynamic cross-sections have been extensively studied \citep{Harding1991, Sina1996, Nobili2008, Gonthier2014, Mushtukov2016}, but they turn out to be cumbersome and the primary bottleneck in radiative transfer codes (see, e.g., \citealt{Araya1999, Schwarm2017, Kumar2022}). To tackle this issue, \cite{Loudas2021} offers approximate, but accurate, resonant magnetic scattering cross-sections that are applicable when $b\ll 1$ and $E \ll m_{\mathrm{e}} c^2$. Given that the systems of our interest lie in this regime, we adopt the resonant Compton scattering prescription developed by \cite{Loudas2021}.

In this study, we neglect polarization effects; therefore, we employ the polarization-averaged differential cross section. In the electron's rest frame, and integrated over the azimuthal angle $\varphi_{\mathrm{rf}}$, it is expressed as
\begin{equation}
\frac{d\sigma}{d\cos\theta'_{\mathrm{rf}}} = 2\pi  \frac{3\pi r_0 c}{16}  L(E_{\mathrm{rf}}, E_{\mathrm{res}})(1+\cos^2\theta_{\mathrm{rf}})(1+\cos^2\theta'_{\mathrm{rf}}),
\label{averaged_cr_sec}
\end{equation}
where $\theta_{\mathrm{rf}}$ and $\theta'_{\mathrm{rf}}$ denote the incident and scattered photon angles relative to the magnetic field, respectively, and $L(E_{\mathrm{rf}},E_{\mathrm{res}})$ is the resonance (Lorentz) line profile,
\begin{equation}
L(E_{\mathrm{rf}},E_{\mathrm{res}})=\dfrac{\Gamma/2\pi}{(E_{\mathrm{rf}}-E_{\mathrm{res}})^2/\hbar^2+(\Gamma/2)^2},
\label{Lorentz_profile}
\end{equation}
centered on the relativistic cyclotron energy, $E_{\mathrm{res}}$ (see Eq.~\ref{res_energy}).
For the relativistic cyclotron transition rate $\Gamma$, we use the first-order corrected expression \citep{Loudas2021} 
\begin{equation}
\Gamma=\dfrac{4}{3}\dfrac{m_{\mathrm{e}} c e^2}{\hbar^2}b^2(1-2.7 b).
\label{Gamma}
\end{equation}
The magnetic scattering cross-section, $\sigma$, is the integral of the differential cross section (\ref{averaged_cr_sec}) over $\theta'_{\mathrm{rf}}$, namely
\begin{equation}
\sigma(E_{\mathrm{rf}},\theta_{\mathrm{rf}})=2\pi\dfrac{\pi r_0c}{2}L(E_{\mathrm{rf}},E_{\mathrm{res}})(1+\cos^2\theta_{\mathrm{rf}}).
\label{integrated_cross_section}
\end{equation}
We employed this expression in the calculation of the effective cross-section \eqref{eq:sigma_eff}.

\subsection{Treatment of the boundary}
\label{sec:boundary_treatment}

Although most photons diffuse away from the base, where they are emitted, and escape, a small fraction of them return to the boundary (base). In this work, we do not model the physics of the NS atmosphere located below the base of the column ($z<0$). So, when a photon packet reaches the base or when the line of sight intersects the base of the cylinder, a special treatment is required. If the wave packet crosses the base at $z=0$, two distinct boundary conditions are considered, "Reflection" or
"Thermalization". 
\begin{itemize}

    \item In the Reflection scheme, the packet undergoes geometric reflection, retaining both its energy and weight. This is analogous to a scattering-dominated, stationary atmosphere. The main body of the paper uses this boundary condition. \\
    
    \item In the Thermalization scheme, the wave packet is instantaneously absorbed and re-emitted isotropically upwards with a new energy sampled from a blackbody distribution (Eq.~\ref{eq:blackbody}), while its weight $w$ remains unchanged. This boundary condition is examined in detail in Appendix \ref{App1}.

\end{itemize}

\section{Results} \label{sec:results}

In this section we present the results of our MC simulations for a typical XRP in the subcritical regime, employing the Reflection scheme (for a comparison with results obtained using the Thermalization scheme, we refer the reader to Appendix \ref{App1}). The seed photon distribution was assumed to be isotropic in the upward hemisphere and uniformly emitted from the base, described by a blackbody spectrum with temperature $T$ given by Eq.~\eqref{eq:temperature}. We adopted a fiducial value of $L^*= 10^{37}\,\mathrm{erg\,s^{-1}}$ for the critical luminosity; the remaining fixed, dimensionless parameters of the model are listed in Table~\ref{tab:model_params}. The local cyclotron energy corresponding to the chosen magnetic field strength at $h=0$ was $E_{\mathrm{c}}=25.55~\mathrm{keV}$. The only remaining free parameter was the relative accretion luminosity, $L/L^*$, which we varied over the range $(0,1)$ in order to study its impact on the emergent spectral features.

\renewcommand{\arraystretch}{1.5}

\begin{table}[h]
    \caption{Fiducial NS parameters adopted in this study.}
    \label{tab:model_params}
    \centering
    \begin{tabular}{l c c}
    \hline\hline
    Parameter & Symbol & Value \\
    \hline 
    Mass & $m$ & $1.4$ \\ 
    Radius & $R_6$ & $1$ \\
    Free-fall speed & $\beta_{\mathrm{ff}}$ & $0.5$ \\ 
   NS magnetic field strength & $b$ & $0.05$ \\ 
   Critical luminosity [$10^{37}~\mathrm{erg~s^{-1}}$] & $L^*$ & $1$ \\
    \hline
    \end{tabular}
\end{table}

In each simulation run, we injected $10^6$ seed photon packets of equal initial weight $w = 10^6$ to ensure high-fidelity spectra. Throughout this work, we present the resulting spectra in the form of $dN/dE$ as a function of the photon energy $E$ and, where indicated, polar angle $\theta$. We normalized the angle-averaged spectra to unity (i.e., $\int (dN/dE) ~dE = 1$) and scaled the angle-resolved ones according to their relative contribution to the total photon flux. 

\subsection{Spectral formation in the accretion channel} \label{spectral_formation}

\begin{figure}[t]
\centering
\includegraphics[width=\hsize]{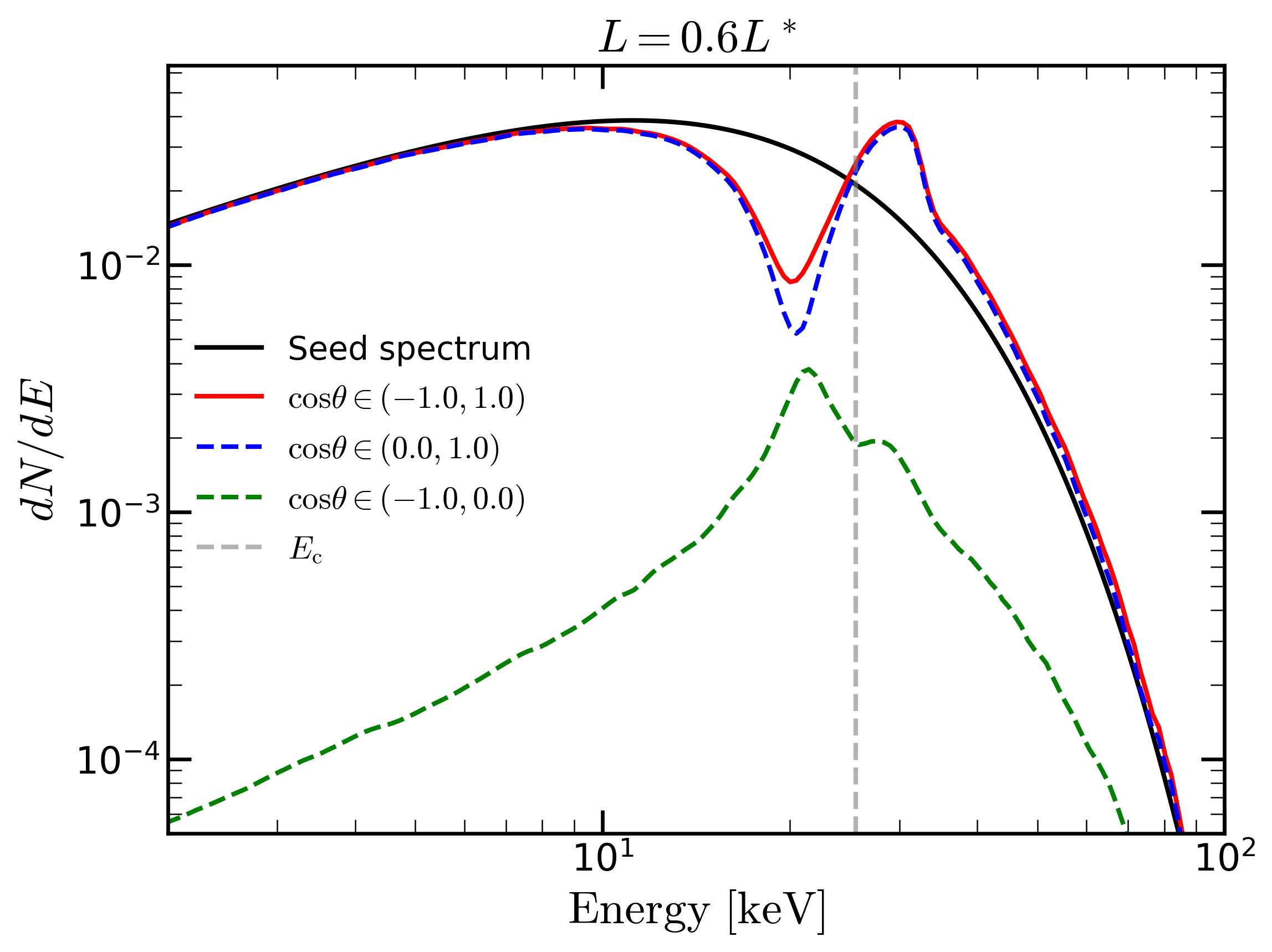}
\caption{Representative emergent spectrum for an XRP with a fiducial luminosity $L=0.6L^*$. 
The solid black line represents the seed blackbody spectrum with temperature $k_{\mathrm{B}} T=5.9~\mathrm{keV}$. 
The solid red line shows the total normalized emergent spectrum. 
The dashed blue and green lines correspond to the angle-selected spectra observed at polar angles in the ranges $(0,\pi/2)$ and $(\pi/2,\pi)$, respectively. 
The vertical dashed gray line indicates the cyclotron energy at $E_{\mathrm{c}}=25.55~\mathrm{keV}$. The spectrum features a redshifted, but prominent CRSF around $E\sim 20\,\mathrm{keV}$  followed by a bump.}
 \label{fig:spectrum}
\end{figure}

To illustrate the primary spectral features and the underlying physical picture, we first performed a simulation for a representative luminosity of $L=0.6\,L^*$. The corresponding emergent spectrum is shown in Fig.~\ref{fig:spectrum}. The solid black line represents the seed blackbody continuum with a temperature of $k_{\mathrm{B}} T=5.9~\mathrm{keV}$. The solid red line shows the total emergent spectrum, while the dashed lines refer to angle-selected emergent spectra: blue (green) denotes photons escaping away from (toward) the NS, with polar angles $\theta \in (0,\pi/2)$ ($(\pi/2,\pi)$). The angle-selected spectra are scaled relative to their contribution to the total photon flux. Finally, the vertical gray line indicates the local classical cyclotron energy adopted in this simulation.

As shown in Fig.~\ref{fig:spectrum}, the resulting spectrum exhibits the expected characteristics outlined in Sect. \ref{sec:CRSF_formation}. The total emergent spectrum closely follows the seed blackbody at energies well below and well above the cyclotron energy, reflecting the low optical depth experienced by photons outside the resonance. The computed spectrum also features a prominent CRSF, resembling an absorption line, followed by a broad blue wing (bump). Photons in resonance with the quantized electrons experience a strongly enhanced opacity and are effectively trapped, undergoing multiple Compton scatterings and gaining on average energy in the process through Doppler boosting until their energies shift out of resonance, resulting in the formation of the CRSF. This energy redistribution of the scattered photons naturally gives rise to the pronounced blue wing (for a comparison with the case where the Thermalization scheme is implemented, see Fig.~\ref{fig:spectrum_new}; there, in contrast, the blue wing is suppressed, with the red wing dominating).

The CRSF appears redshifted with respect to the local cyclotron energy. The origin of this shift is related to the Doppler effect between the lab frame and the electron's bulk rest frame, as described qualitatively in Sect. \ref{sec:CRSF_formation}. This effect is further discussed in more detail in the following paragraphs. 

Furthermore, we note that the dominant contribution to the total flux arises from photons escaping at angles in the range $(0,\pi/2)$, owing to the low optical depth values away from resonance and the reflection (or thermalization) of photons crossing the base of the funnel. Yet, a minor fraction of photons escape towards the NS surface. The spectrum of this latter population does not exhibit a prominent absorption feature at the energies corresponding to the CRSF in the total spectrum, since photons propagating with $\cos\theta < 0$ are typically out of resonance at those energies. Instead, an absorption feature appears at higher energies, where photons are in resonance with the electrons, but it is strongly smeared out by the contribution of red- and blue-wing photons propagating in different directions. Escaped photons directed towards the NS will likely be reflected off the surface; however, their contribution to the total spectrum is negligible; We therefore safely neglect this photon population and the associated modeling of surface reflection outside the funnel. 

\begin{figure}[t]
\centering
\includegraphics[width=\hsize]{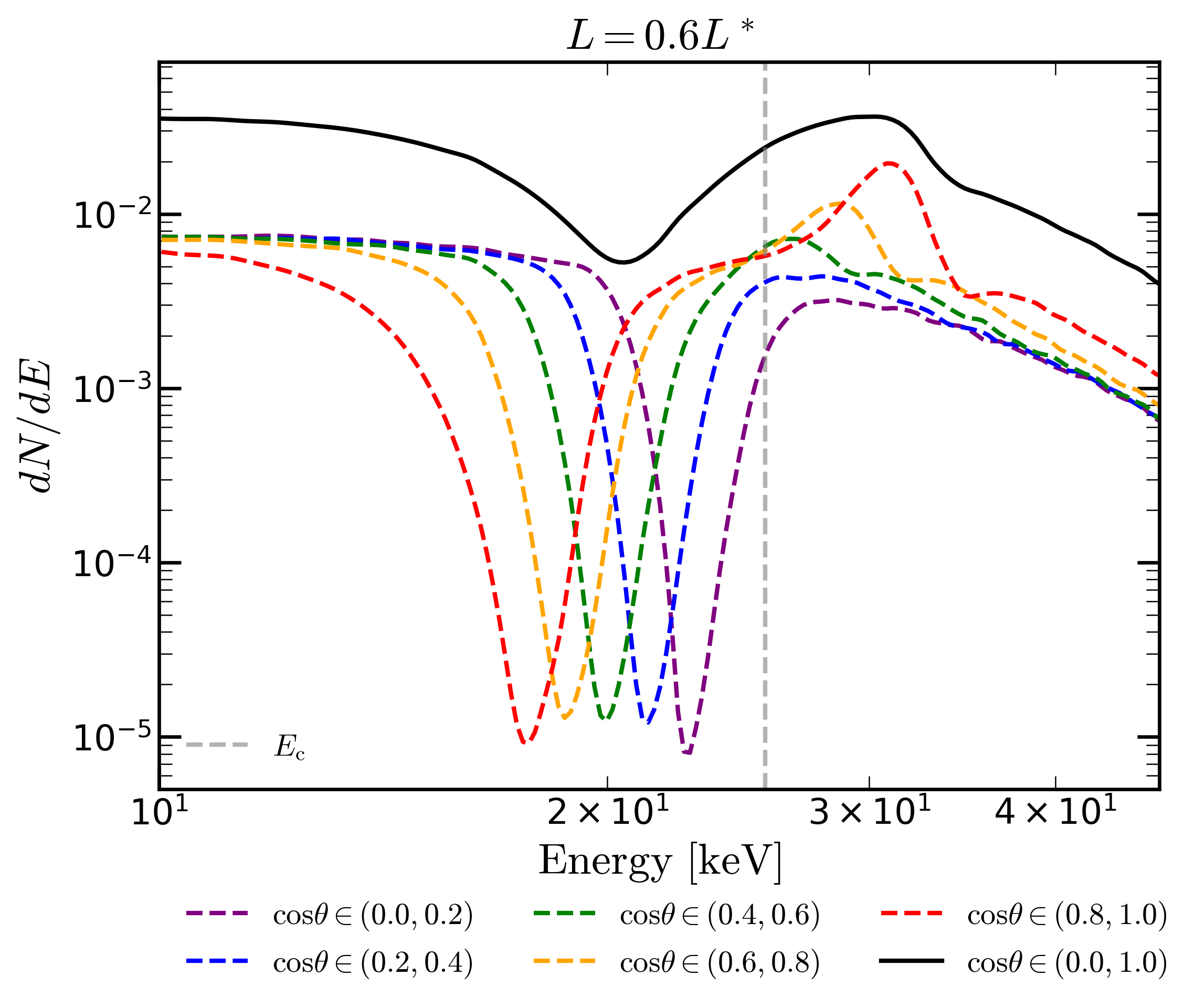}
\caption{Angle-resolved spectra for a typical XRP source with luminosity $L=0.6L^*$. The solid black line corresponds to the angle-averaged spectrum in the range $(0, \pi/2)$. The dashed colored lines show spectra extracted in five angular bins within this range. Each spectrum is normalized to its angular flux contribution. The vertical gray line marks the local cyclotron energy. With increasing $\cos\theta$, the CRSF centroid energy shifts to lower values and the blue wing becomes more pronounced.}
\label{fig:angle_resolved}
\end{figure}

To further illuminate the formation of the CRSF, we investigated the angular dependence of the emergent spectrum, which is relevant for phase-resolved studies of CRSFs. Fig.~\ref{fig:angle_resolved} shows the total spectrum in the range $(0, \pi/2)$ in black, along with decomposed into five narrow viewing angle bins spectra in various colors. We note that the choice of the number of angular bins was arbitrary and does not affect the conclusions. The angle-dependent spectra were scaled based on their relative contribution in the photon flux. For convenience, the position of the classical cyclotron energy is also marked by a dashed vertical gray line.

Evidently, the CRSF profile and its wings depend strongly on the viewing angle, exhibiting two primary characteristics. First, the centroid energy of the CRSF decreases with increasing $\cos\theta$. Specifically, as $\cos\theta$ approaches zero (corresponding to directions perpendicular to the $z$-axis), the redshift relative to $E_{\mathrm{c}}$ becomes smaller. This behavior arises from the Doppler effect discussed in Sect. \ref{sec:CRSF_formation}. A photon propagating at an angle $\theta$ with respect to the $z$-axis is in resonance with the infalling electrons when its energy is approximately $E_{\mathrm{c}} /\gamma (1 + \beta \cos \theta)$ (see Eq.~\ref{redshift_energy}); thus, larger values of $\cos \theta$ correspond to lower resonant energies, naturally producing the trend shown in Fig.~\ref{fig:angle_resolved}. 

Second, the blue wing becomes increasingly prominent and shifts to higher energies as $\cos \theta$ increases. This behavior is a direct consequence of the implementation of the Reflection scheme: seed photons undergoing resonant scattering acquire their maximum energies when they are backscattered, and therefore Doppler beamed toward the stellar surface (i.e., at angles near $\cos\theta \approx -1$). Upon reflection from the base of the funnel, these high-energy photons re-emerge at viewing angles near $\cos \theta \approx 1$, manifesting as prominent blue wings.

Finally, we find that the angle-averaged spectrum features a broader, asymmetric, but shallower CRSF compared to those in angle-resolved spectra. The superposition of CRSFs formed at different viewing angles--each centered at a slightly different energy--produces a wider and shallower feature due to the partial overlap of individual absorption profiles with the wings of neighboring lines. At low luminosity, this is very prominent (see Appendix \ref{App3}.).

\begin{figure}[t]
\centering
\includegraphics[width=\hsize]{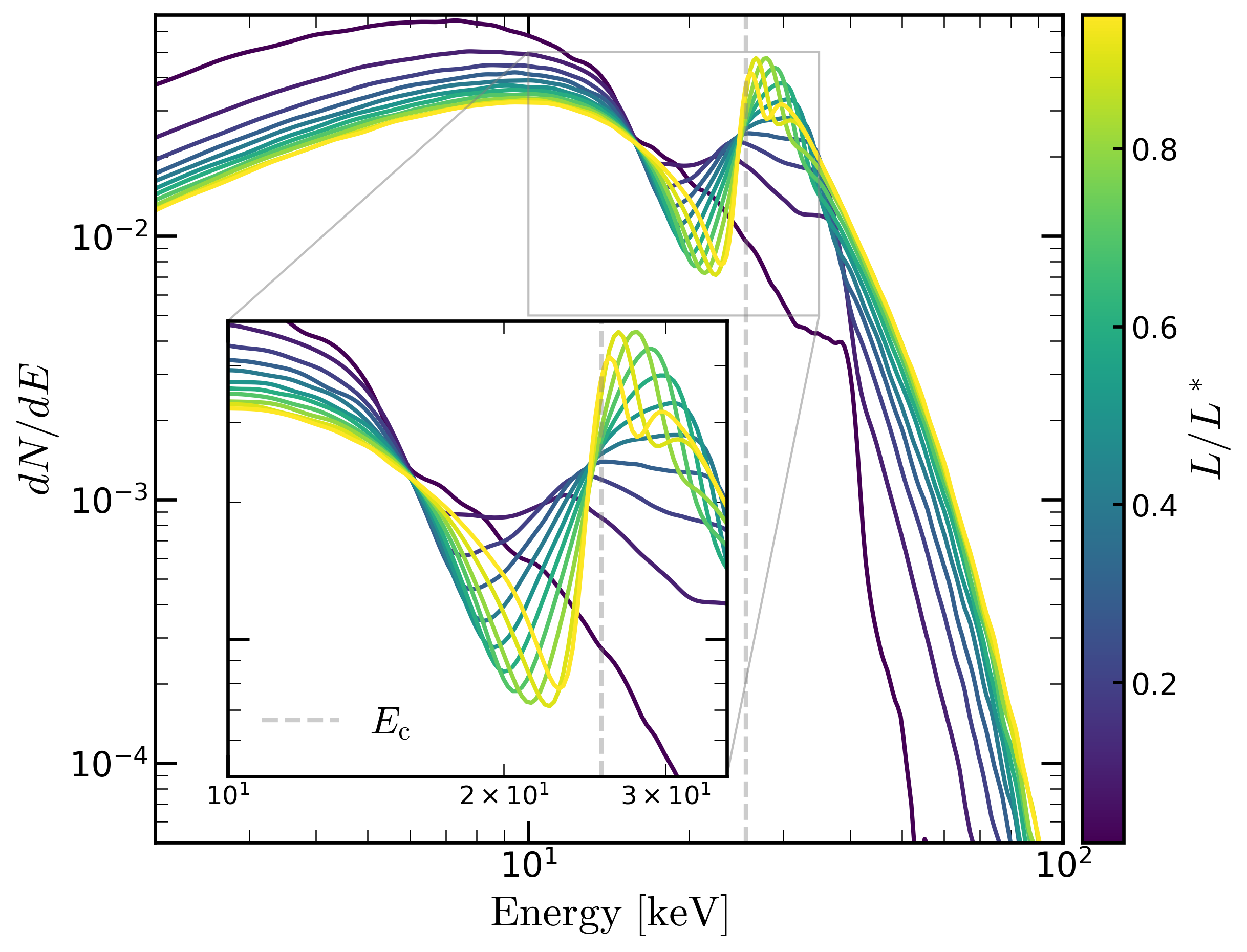}
\caption{Angle-averaged emergent spectra for luminosities spanning the range $L/L^*\in (0,1)$. The color gradient from dark blue to yellow corresponds to increasing luminosity. The inset highlights the spectral region around the CRSF. The vertical gray dashed line denotes the local cyclotron energy. The positive correlation between $L$ and the CRSF energy $E_{\mathrm{CRSF}}$ is evident.}
 \label{fig2}
\end{figure}

\subsection{Variability due to the accretion luminosity}
\label{sec:variability_luminosity}

Having studied the angular dependence of the CRSF properties and the physical origin of the associated spectral features, we now turn to their dependence on the accretion luminosity. At higher luminosities, the braking of the infalling plasma becomes more efficient; consequently, the Doppler effect, which causes the cyclotron energy to appear redshifted, is reduced. As a result, the centroid energy is expected to lie closer to the local value $E_{\mathrm{c}}$ at higher luminosities (see also Sect. \ref{sec:CRSF_formation}).

We performed a total of 11 simulations, varying the accretion luminosity while keeping all other model parameters fixed. Figure \ref{fig2} displays the angle-averaged emergent spectra for different luminosities, with the inset highlighting the energy range around the CRSF. All spectra are normalized to unity. As anticipated, the CRSF centroid energy shifts systematically toward higher energies with increasing luminosity, establishing a positive correlation between $L$ and $E_{\mathrm{CRSF}}$ (for a quantitative characterization, see Sect. \ref{sec:phase-resolved}), in agreement with the predictions of \cite{Mushtukov2015b}. 

In addition, the blue wing becomes increasingly pronounced and the CRSF deepens with increasing luminosity. The overall spectrum also hardens. Higher luminosities correspond to larger plasma number densities (and thus optical depth), as Eq.~\eqref{eq:density_profile} suggests, due to (i) an increased accretion rate and (ii) enhanced compression associated with a reduced terminal velocity. The resulting increase in optical depth naturally accounts for the deepening of the CRSF, the strengthening of the blue wing, and the observed spectral hardening.

\begin{figure*}[ht] 
    \centering

    \begin{subfigure}[b]{0.49\textwidth}
        \centering
    \includegraphics[width=\linewidth]{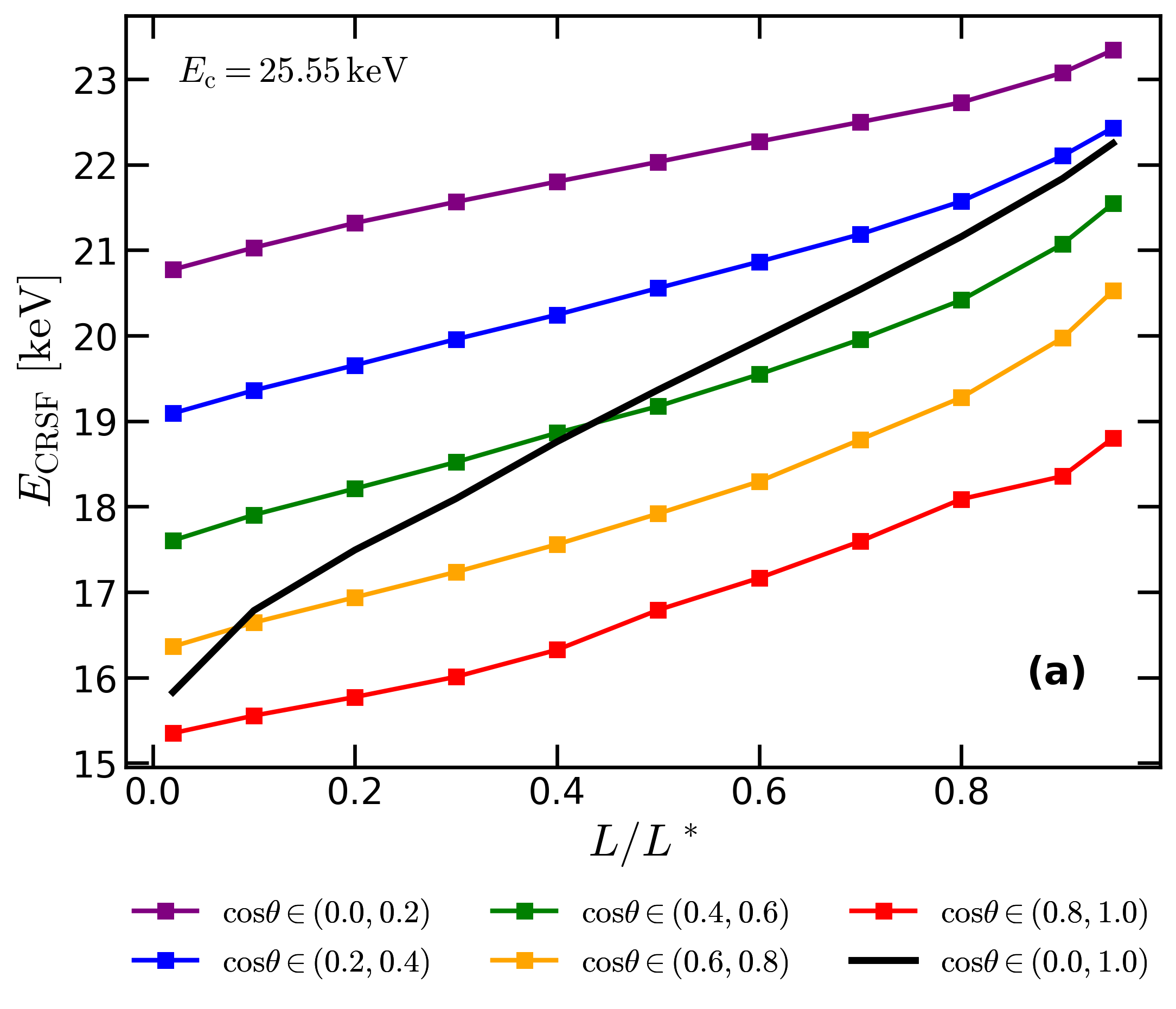}%
      \phantomcaption
        \label{fig:E_vs_L}
    \end{subfigure}
    \hfill 
    \begin{subfigure}[b]{0.49\textwidth}
        \centering
    \includegraphics[width=\linewidth]{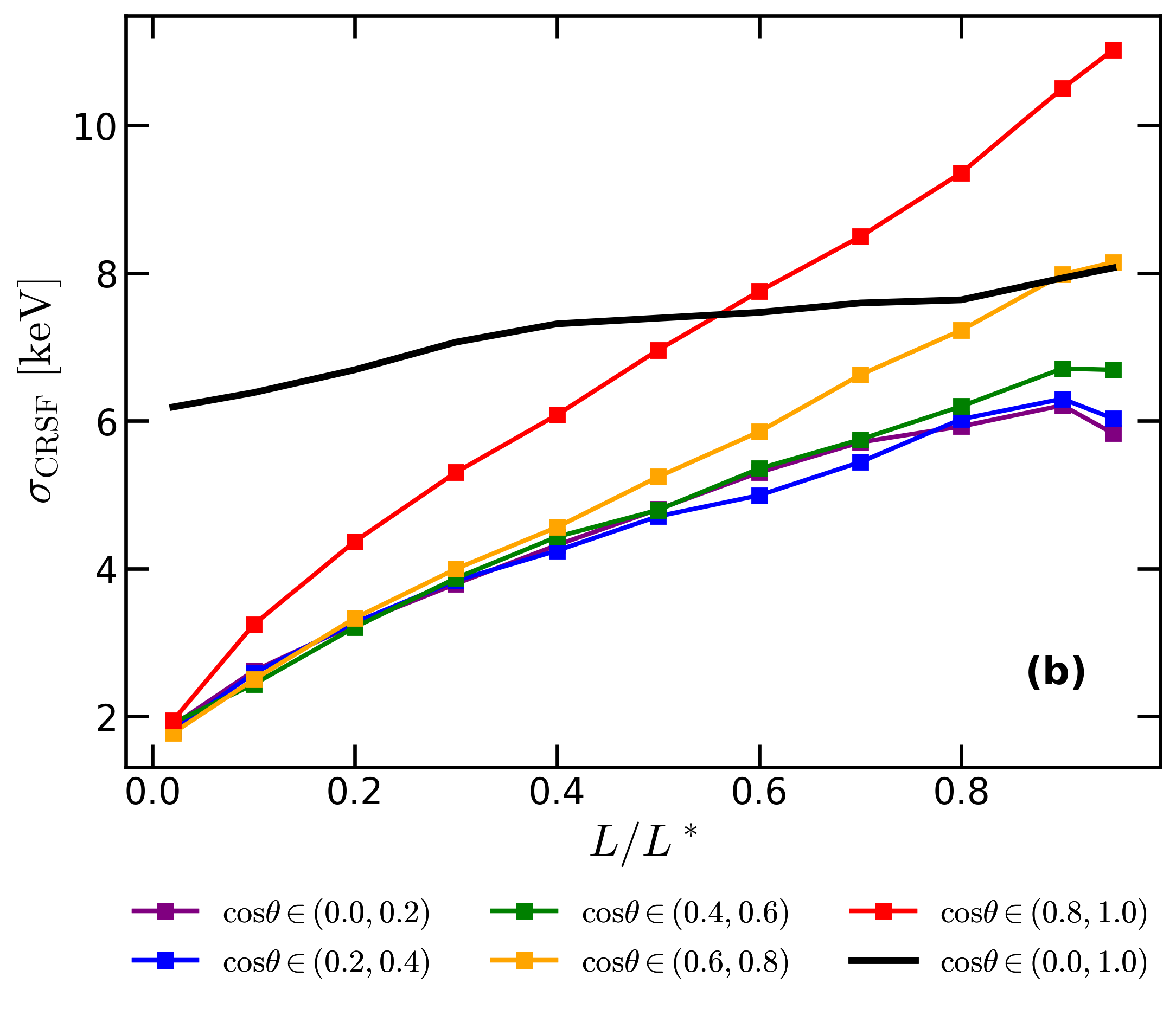}%
       \phantomcaption
        \label{fig:width_vs_L}
    \end{subfigure}
    \caption{Dependence of the CRSF centroid energy $E_{\mathrm{CRSF}}$ (panel a) and width $\sigma_{\mathrm{CRSF}}$ (panel b) on luminosity and viewing angle $\theta$ (measured with respect to the magnetic axis). Results are shown for five angular bins spanning the range $\theta \in (0,\pi/2)$, with $E_{\mathrm{c}} = 25.55\,\mathrm{keV}$. Solid black lines denote quantities extracted from angle-averaged spectra. Both $E_{\rm CRSF}$ and $\sigma_{\rm CRSF}$ exhibit a positive correlation with luminosity for all viewing angles. At fixed luminosity, $E_{\mathrm{CRSF}}$ ($\sigma_{\mathrm{CRSF}}$) decreases (increases) toward smaller viewing angles, indicating substantial variability over the pulse cycle.}
    \label{fig:combined_horizontal}
\end{figure*}

\subsection{Phase-dependent variability of the line features}
\label{sec:phase-resolved}

Having demonstrated the strong variability of the CRSF with luminosity (Sect. \ref{sec:variability_luminosity}), and viewing angle (Sect. \ref{spectral_formation}), we now quantify the dependence of the line centroid energy $E_{\rm CRSF}$ and width $\sigma_{\rm CRSF}$ on these two parameters. Figure \ref{fig:E_vs_L} displays the CRSF centroid energy as a function of luminosity for five angular intervals spanning the range $(0, \pi/2)$, together with the values measured from the angle-averaged spectra in black. We defined $E_{\mathrm{CRSF}}$ as the centroid of the CRSF after subtracting the underlying continuum (see Appendix \ref{App2}).
A clear positive correlation between $E_{\rm CRSF}$ and luminosity is present for all viewing angles, as well as for angle-averaged spectra. 

 At fixed luminosity, the centroid energy varies systematically with viewing angle: smaller angles relative to the $z$-axis (i.e., face-on views of the hot spot) consistently yield lower values of $E_{\mathrm{CRSF}}$, indicating an anticorrelation with $\cos\theta$. An explanation is given in Sect. \ref{spectral_formation}. The amplitude of this angular variation is $\sim 30\%$ at low luminosities and decreases as the luminosity increases. From Eq.~\eqref{redshift_energy}, the resonant energy range of photons moving away from the star spans $[E_{\rm CRSF}(\theta = 0), E_{\rm CRSF}(\theta = \pi/2)]$, that is
\begin{equation}
    \dfrac{\Delta E|_{\rm CRSF}}{E_{\rm res}} \approx \dfrac{1}{\gamma} - \dfrac{1}{\gamma( 1 + \beta)} = [\gamma (1 + 1/\beta)]^{-1}.
\end{equation}
Substituting $\beta$ with the terminal velocity $\beta_0$, we get $\Delta E|_{\rm CRSF}/E_{\rm res} \approx 29\%$ for $\beta_0 = 0.5$, which corresponds to $L\ll L^*$. At higher luminosities, $\beta_0$ decreases (Eq.~\ref{eq:beta_0}), leading to a smaller $\Delta E|_{\rm CRSF}$, in agreement with the simulation results.  

Figure \ref{fig:width_vs_L} is analogous to Fig.~\ref{fig:E_vs_L}, but shows the CRSF width. While several definitions of line width exist, depending on the assumed profile, here we characterized $\sigma_{\mathrm{CRSF}}$ as the full width at half minimum (FWHM) of the CRSF after subtracting the underlying continuum; for details we refer the reader to Appendix \ref{App2}. We find that the line width increases with luminosity for all viewing angles, although its absolute value depends substantially on $\theta$. The widths corresponding to the angle-averaged spectra (black curve) are systematically higher than those in angle-dependent spectra, as the width in the former case is dominated by the superposition of CRSFs forming at different angles (see Figs.~\ref{fig:angle_resolved} \& \ref{fig:angle_resolved_overlap}). In addition, we find that $\sigma_{\mathrm{CRSF}}$ is positively correlated with $\cos\theta$, in agreement with the predictions of \cite{Mushtukov2015b}. 

This behavior can be easily understood qualitatively as follows. Higher luminosities produce stronger radiative drag, resulting in a steeper velocity gradient in the accretion flow (see Eq.~\ref{eq:velocity_profile}). Photons traversing the funnel therefore sample a broader range of electron bulk velocities, leading to increased Doppler broadening of the CRSF and a positive correlation between $\sigma_{\mathrm{CRSF}}$ and $L$. For a given velocity profile (i.e., fixed luminosity), the magnitude of this effect depends on the propagation direction relative to the flow (in this case, the $z$-axis): smaller values of $\cos\theta$ suppress Doppler shifts (see Eq.~\ref{redshift_energy}), while the maximum broadening occurs for photons propagating nearly parallel to the magnetic (flow) axis, $\cos\theta \approx 1$. 

Taken together, these results imply that both the CRSF centroid energy and width increase with luminosity, such that higher values of $E_{\mathrm{CRSF}}$ are accompanied by broader lines.
Consequently, in studies of phase-averaged spectral evolution, this model predicts a correlated increase of the CRSF energy and width with luminosity. 
Conversely, the trend reverses for short-timescale, phase-resolved variability. As shown above, $E_{\mathrm{CRSF}}$ anticorrelates with $\cos\theta$, whereas $\sigma_{\mathrm{CRSF}}$ correlates positively with $\cos\theta$. For a given luminosity, therefore, yields an anticorrelation between the CRSF energy and width. Over a pulse cycle, the line energy (width) reaches its minimum (maximum) value when the accretion funnel is viewed face-on, i.e., along the magnetic-field direction, and its maximum (minimum) when it is seen edge-on.

 \section{Comparison with data: an application to the X-ray pulsar GX 304$-$1} \label{sec:comparison}

Following \cite{Mushtukov2015b}, we tested our model by applying it to the observed CRSF variability in the X-ray pulsar GX 304$-$1, a source exhibiting well-established $E_{\mathrm{CRSF}}-L$ and $\sigma_{\mathrm{CRSF}}-L$ positive correlations \citep{Klochkov2012}. We adopted the CRSF centroid energy and width measurements reported by \cite{Mushtukov2015b}, who analyzed INTEGRAL \citep{Winkler2003} observations of GX 304$-$1 obtained during the 2012 type-II outburst \citep{Yamamoto2012}.

To facilitate a direct comparison with the data, we performed an additional set of MC simulations employing the Reflection scheme, as described in Sect. \ref{sec:results}, but, following \cite{Mushtukov2015b}, we took the classical cyclotron energy to be $E^{GX}_c=66.5\,\mathrm{keV}$, corresponding to a dimensionless magnetic field strength of $b \approx 0.13$, and adopted a critical luminosity of $L^*_{GX}=2.7\times 10^{37}\,\mathrm{erg\,s^{-1}}$. The remaining dimensionless NS parameters used in this exercise are listed in Table~\ref{tab:GX_params}. The photon injection protocol was identical to that utilized in runs shown in Sect. \ref{sec:results}.

\begin{figure*}[t]
    \centering
    \begin{subfigure}[b]{0.49\textwidth} 
        \centering
    \includegraphics[width=\linewidth]{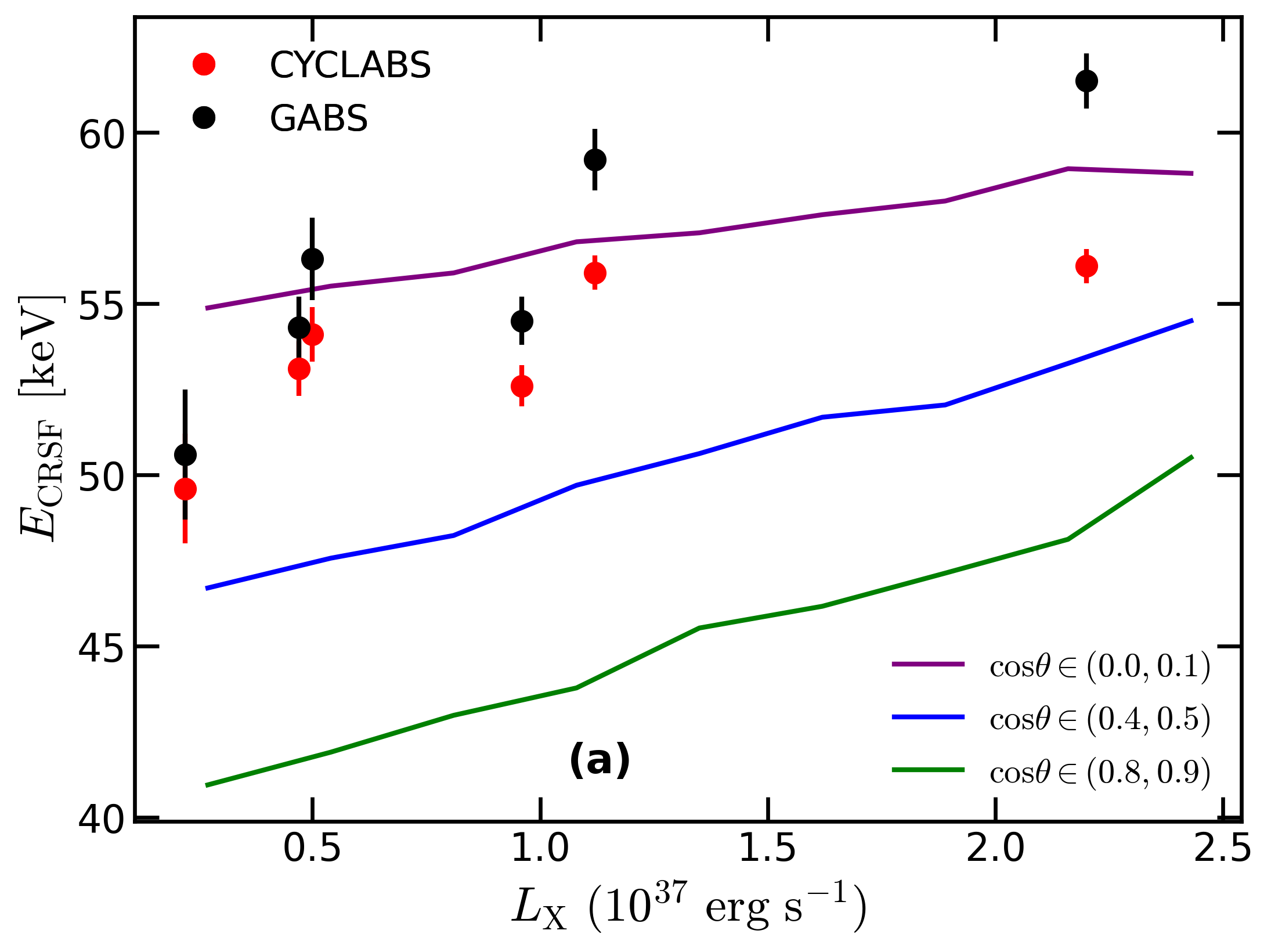}%
    \phantomcaption
        \label{fig:E_vs_L_rel}
    \end{subfigure}
    \hfill 
    \begin{subfigure}[b]{0.505\textwidth}
        \centering
    \includegraphics[width=\linewidth]{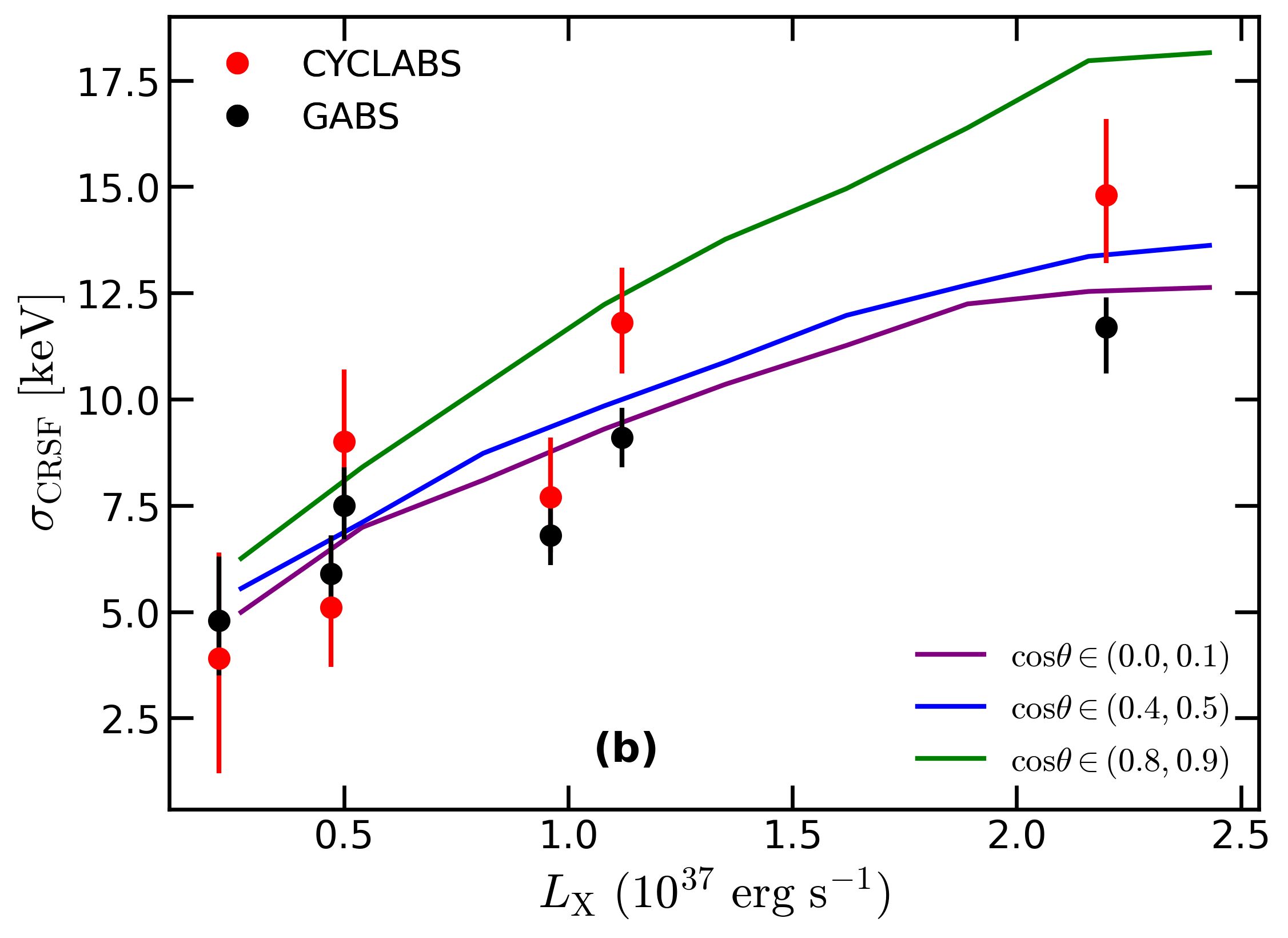}%
    \phantomcaption
        \label{fig:width_vs_L_real}
    \end{subfigure}

    \caption{Comparison of observed cyclotron line variability in GX 304-1 with our model. Panel a (b) shows the CRSF centroid energy (width). Red and black circles represent the observed $E_{\mathrm{CRSF}}$ and $\sigma_{\mathrm{CRSF}}$ values of GX 304-1, obtained using the CYCLABS model (Lorentzian profile) and the GABS model (Gaussian profile), respectively, to fit the line feature \citep{Mushtukov2015b}. Green, blue, and purple lines correspond to the results derived from our MC simulations for three different viewing angle intervals. The model predictions, for specific viewing angles, agree sufficiently well with the data over an order of magnitude in luminosity.}
    \label{fig:combined_plots}
\end{figure*}

Figures~\ref{fig:E_vs_L_rel} and \ref{fig:width_vs_L_real} show the GX 304$-$1 data (red and black circles) together with our simulation results (colored lines) for $E_{\mathrm{CRSF}}$ (left panel) and $\sigma_{\mathrm{CRSF}}$ (right panel) as functions of luminosity, for three representative narrow viewing-angle intervals. We measured $E_{\mathrm{CRSF}}$ and $\sigma_{\mathrm{CRSF}}$ from the simulated spectra as described in Sect. \ref{sec:phase-resolved}. Regarding the data, black circles correspond to CRSF parameters inferred using a Gaussian absorption profile (GABS), while red circles denote results obtained using a Lorentzian profile (CYCLABS). Further details concerning the data analysis can be found in Sect. 3 of \cite{Mushtukov2015b}.

The model successfully reproduces the observed trends and, for fixed viewing angles, provides good agreement with the data over nearly an order of magnitude in luminosity. In particular, viewing angles corresponding to $\cos \theta \in (0.0-0.1)$ (purple curves) simultaneously match both the CRSF centroid energy and width, whereas  larger values of $\cos\theta$ are inconsistent with one or both observables. Since the viewing angle is primarily governed by the system geometry (neglecting gravitational light bending, which is not included here), this result favors configurations in which the accretion funnel (and thus the hot spot) is viewed predominantly edge-on over the pulse cycle. Such configurations may correspond either to a near-orthogonal rotator ($\alpha \sim 70^\circ - 90^\circ$) observed at low inclination ($i \sim 0^\circ - 20^\circ$), or to a near-aligned rotator ($\alpha \sim 0^\circ-20^\circ$) observed at high inclination ($i \sim 70^\circ - 90^\circ$). Here, $i$ denotes the angle between the line of sight and the rotation axis, and $\alpha$ is the inclination of the magnetic axis relative to the rotation axis. The latter scenario is the one proposed by \cite{Mushtukov2015b}, based on phase-averaged modeling of the CRSF variability, and is consistent with our findings.

\begin{table}
    \caption{GX 304-1 model parameters.}
    \label{tab:GX_params}
    \centering
    \begin{tabular}{l c c}
    \hline\hline 
    
    Parameter & Symbol & Value \\
    \hline
    
    Mass & $m^{GX}$ & $1.4$ \\ 
    
    Radius & $R_{6}^{GX}$ & $1$ \\ 
    
    Free-fall speed & $\beta_{\mathrm{ff}}^{GX}$ & $0.5$ \\ 
   NS magnetic field strength & $b$ & $0.13$ \\
   
   Critical luminosity $[10^{37}~\mathrm{erg~s^{-1}}]$ & $L^*_{GX}$ & $2.7$ \\
    \hline
    \end{tabular}
\end{table}

Nevertheless, it would be premature to use this comparison to place quantitative constraints on the physical parameters of GX 304$-$1. Our analysis does not include phase averaging over the pulse cycle, nor does it account for general-relativistic light bending. Additional limitations arise from parameter degeneracies inherent to our modeling approach, including the assumed values of $E^{GX}_c$, $L^*_{GX}$, and other NS parameters. For instance, adopting a higher value of $E^{GX}_c$ would favor larger values of $\cos\theta$ in order to maximize the Doppler redshift, potentially altering the inferred geometry. Likewise, increasing $\beta_{\mathrm{ff}}$ would steepen the predicted $E_{\mathrm{CRSF}}-L$ correlation. A comprehensive comparison with the data therefore requires treating all relevant model parameters as free and performing full phase-averaged calculations, which we defer to future work.

 \section{Discussion \& Conclusions} \label{sec:discussion}

In this work, we investigated quantitatively the origin of the CRSF variability observed in subcritical ($L \lesssim L^*$) XRPs by means of resonant scattering in the accretion funnel above the hot spot, following the model proposed by \cite{Mushtukov2015b}. We focused on two key spectral properties: the positive correlations of the CRSF centroid energy, $E_{\rm CRSF}$, and line width, $\sigma_{\rm CRSF}$, with the X-ray luminosity, $L$. 

We developed a relativistic MC code, based on the framework described by \cite{Loudas2023}, to perform detailed radiative-transfer calculations in the accretion funnel and derive emergent, angle-dependent spectra. For simplicity, we assumed a semi-infinite cylindrical accretion channel above the hot spot, centered on the magnetic axis. Seed blackbody radiation was injected isotropically upward from the hot spot. The accretion flow was taken to be cold, while the plasma density and velocity profiles were computed analytically, capturing the effects of radiation pressure.

The primary model parameters governing our calculations were the critical luminosity $L^*$ and the surface magnetic field strength $B$. Other free parameters - such as the NS mass, NS radius, and free-fall speed - were fixed to fiducial values. We varied the accretion luminosity $L$ over a wide range to study its impact on the CRSF properties. As a proof-of-concept, we applied our model to observational data of the X-ray pulsar GX 304$-$1.

Our main conclusions can be summarized as follows:

\begin{enumerate}

    \item Resonant scattering of hot spot radiation by the accretion flow imprints a prominent CRSF, followed by a broad blue wing (Fig.~\ref{fig1}). The CRSF is always redshifted relative to the local cyclotron energy, $E_{\mathrm{c}}$, due to the Doppler effect; the magnitude of the redshift decreases both at higher luminosities (Fig.~\ref{fig2}) and at larger viewing angles relative to the magnetic axis (Fig.~\ref{fig:angle_resolved}).

    \item The CRSF centroid energy and width are positively correlated with luminosity for all viewing angles (Fig.~\ref{fig:combined_horizontal}), reflecting changes in the accretion-flow velocity profile, although their absolute values vary across viewing angles. With increasing luminosity, the blue wing becomes increasingly pronounced, the CRSF deepens, and the overall spectrum hardens due to the increase in the optical depth (Fig.~\ref{fig2}). 

    \item The CRSF properties exhibit strong variability over the pulse cycle. At fixed luminosity, $E_{\rm CRSF}$ ($\sigma_{\rm CRSF}$) decreases (increases) with $\cos\theta$; the centroid energy reaches its maximum when the funnel is viewed edge-on due to the suppression of the Doppler shift. In contrast, the line width is maximized when the funnel is viewed face-on, as photons sample a broader range of bulk velocities. Thus, in phase-resolved studies, $E_{\rm CRSF}$ anticorrelates with $\sigma_{\rm CRSF}$ during the pulse cycle (Fig.~\ref{fig:combined_horizontal}).

    \item Angle-averaged spectra feature broader, asymmetric, but shallower CRSFs compared to angle-resolved spectra (i.e., at a given pulse phase), owing to the superposition of CRSFs formed at different viewing angles (Fig.~\ref{fig:angle_resolved}).

    \item When applied to the observational data of the X-ray pulsar GX 304$-$1, the model successfully reproduces the observed variability of the CRSF centroid energy and width over nearly an order of magnitude in luminosity (Fig.~\ref{fig:combined_plots}). In particular, configurations in which the accretion channel is predominantly viewed edge-on are favored, in line with \cite{Mushtukov2015b}, providing tentative constraints on the viewing geometry of the system (see Sect. \ref{sec:comparison}).
    
\end{enumerate}

Our findings are consistent with the qualitative predictions of \cite{Mushtukov2015b}, supporting the hypothesis that the accretion channel constitutes the line-forming region and that the flow velocity profile drives the observed variability. However, unlike those predictions, which rely solely on the behavior of the resonant optical depth, our calculations self-consistently incorporate key radiative-transfer effects, such as photon energy redistribution during scattering, which is responsible for the formation of the CRSF wings. 

While the model captures the essential physics governing CRSF variability in subcritical XRPs and successfully reproduces well-established observational trends, it relies on a number of simplifying assumptions that introduce limitations and motivate future extensions. First, we assumed a simplified cylindrical, filled accretion channel; however, more realistic magnetospheric flows point to bow-shaped or annular footprints \citep[see e.g.,][and references therein]{Zhu2025}, which nevertheless are not expected to alter our conclusions. Second, the funnel was extended to infinity, neglecting the global magnetospheric accretion picture, and assuming a constant magnetic field strength. Neither assumption is physical, however, as we argued in Sect. \ref{sec:CRSF_formation}, the accretion flow is optically thick to resonant scattering; therefore, the line-forming region lies above the hot spot and is relatively compact. Considering a dipole field ($B(r) \propto r^{-3}$) and a vertical size of the line-forming region comparable to the hot spot size (Eq.~\ref{eq:hot spot_size}), the expected variation due to the magnetic field gradient is $\lesssim (3-6)~ \%$.

We employed analytical, one-dimensional velocity and density profiles, however, RMHD simulations \citep{Markozov2023} reveal more complex profiles, as the effect of radiation pressure force is less effective near the edges of the funnel, where radiation can easily escape sideways. This subtlety might affect the exact CRSF shape and slope of predicted correlations, but it would not change our conclusions. In addition, we treated the accretion flow as cold (i.e., $k_{\mathrm{B}}T_{\mathrm{e}} \ll E_{\mathrm{c}}$). The electron's temperature is expected to modify the CRSF shape through thermal broadening. So, this approximation is valid, so long as the Doppler effect and the velocity gradient govern the CRSF width. In Sect. \ref{sec:results}, we showed that for a typical cyclotron line energy of $E_{\mathrm{c}} \approx 25\,\mathrm{keV}$, $\sigma_{\mathrm{CRSF}}$ is always greater than $2\,\mathrm{keV}$. Thus for electron temperatures of $k_{\mathrm{B}}T_{\mathrm{e}} \lesssim 2\,\mathrm{keV}$, thermal effects will likely not be important.

Our model does not include a detailed description of the NS atmosphere beneath the accretion funnel (i.e., $z < 0$) or the associated physical processes. To assess the sensitivity of our results to the treatment of the lower boundary, we considered two limiting cases: one in which photons reaching the base undergo geometric reflection (Reflection scheme), conserving their energy, and another in which photons get absorbed and re-emitted as blackbody radiation (Thermalization scheme). We find that our primary conclusions regarding the CRSF variability are robust against the choice of boundary condition. Nevertheless, a more physically motivated treatment that self-consistently accounts for the structure, plasma dynamics, and radiative transfer in the NS atmosphere would further strengthen the predictive power of the model. Such an extension is beyond the scope of the present work and is left for future explorations.

Another simplification in our calculations concerns the treatment of the hot spot emission. We assumed isotropic, upward blackbody emission, uniformly from the base of the accretion funnel. In contrast, more detailed hot spot emission models predict anisotropic, Comptonized spectra \citep[see e.g.,][]{Meszaros1985, Sokolova-Lapa}. To assess the sensitivity of our results to this assumption, we tested alternative emission prescriptions, including pencil-like patterns and point-source emission centered at the origin, $(0,0,0)$. In all cases, the resulting variations in the CRSF centroid energy and width were at $\lesssim 10\%$ level. This indicates that the CRSF variability explored in this work is not strongly sensitive to the assumed hot spot emission pattern, although such effects are expected to play a critical role in modeling of pulse profiles, which will be explored in future work.

In this work, we performed polarization-independent calculations. With the current (IXPE \citealt{IXPE2022}; XL-Calibur \citealt{XLCALIBUR2021}; XPoSat\footnote{\url{https://www.isro.gov.in/XPoSat.html}}) and upcoming (eXTP \citealt{eXTP2025}) X-ray polarimetry missions, polarization data are becoming increasingly available. X-ray polarization holds strong constraining power regarding the system's geometry \citep{Poutanen2024}; thus, incorporating polarization into our radiative transfer scheme is a vital next step. Our forthcoming work is expected to address this topic in detail.

Finally, we neglected general relativistic effects in this analysis, which can be important for pulse-profile modeling and thorough comparisons with observations \citep[see e.g.,][]{Markozov2024}. Such effects may influence our results through: (i) gravitational redshift of the CRSF energy and width, as well as the inferred luminosity, and (ii) gravitational lensing by the NS. While gravitational redshift may shift the absolute values of the CRSF parameters, this shift is expected to be nearly identical in all cases, since the extent of the line-forming region above the hot spot is small. Thus, it should not affect the variability of the CRSF dramatically. Similarly, gravitational light bending will modify the observed range of viewing angles and introduce contributions from the antipodal hot spot. These effects are expected to become important when constructing the predicted pulse profiles. 

An alternative interpretation of CRSF variability in subcritical XRPs is provided by the CS scenario, which identifies the line formation site as the region between the hot spot and the shock front. In this framework, the observed variations in the CRSF energy and width are attributed to changes in the local magnetic field strength at the line-forming region, which are tied to the shock altitude. When applied to the GX 304$-$1 system, this model successfully reproduces the luminosity dependence of the CRSF properties in phase-averaged spectra \citep{Rothschild2017}. Because bulk motion effects in the post-shock region are negligible, pulse-resolved CRSFs are not expected to exhibit strong variability with the pulse phase (i.e., viewing angle), in contrast to the scenario explored in this work. This fundamental difference between the two models offers a potential observational discriminant that could help distinguish between competing interpretations and advance our understanding of cyclotron line formation in subcritical XRPs. However, such a test requires the development of a robust model for the  shock structure, followed by self-consistent radiative-transfer calculations.

In summary, building on the framework proposed by \cite{Mushtukov2015b}, we have quantitatively demonstrated the physical origin of key observational trends in subcritical X-ray pulsars, including the luminosity- and phase-dependent variability of cyclotron lines, by means of resonant scattering in the accretion funnel. Our results highlight the central role of bulk motion in shaping the CRSF properties and provide new insights into the dynamics and geometry of the accretion flow.

\section*{Data availability}

The results presented in this paper were obtained using a
proprietary code developed by the authors. The data supporting the figures, as well as the code, are available from the corresponding author upon reasonable request.
 
\begin{acknowledgements}
    PF thanks Chryssi Koukouraki and Orestis Zoumpoulakis (Institute of Astrophysics, Foundation for Research and Technology-Hellas) for their technical assistance.
    NL acknowledges support from the Venture Forward Fellowship of the Department of Astrophysical Sciences at Princeton University. 
    NL thanks Lizhong Zhang for valuable discussions on accretion column dynamics, and Anatoly Spitkovsky for insightful discussions on the underlying physics of the collisionless shock scenario as an accretion-halting mechanism above the hot spot.
\end{acknowledgements}

\bibliography{references.bib}

\begin{appendix}

\section{Boundary effects}
\label{App1}

\begin{figure}[t]
\centering
\includegraphics[width=0.95\hsize]{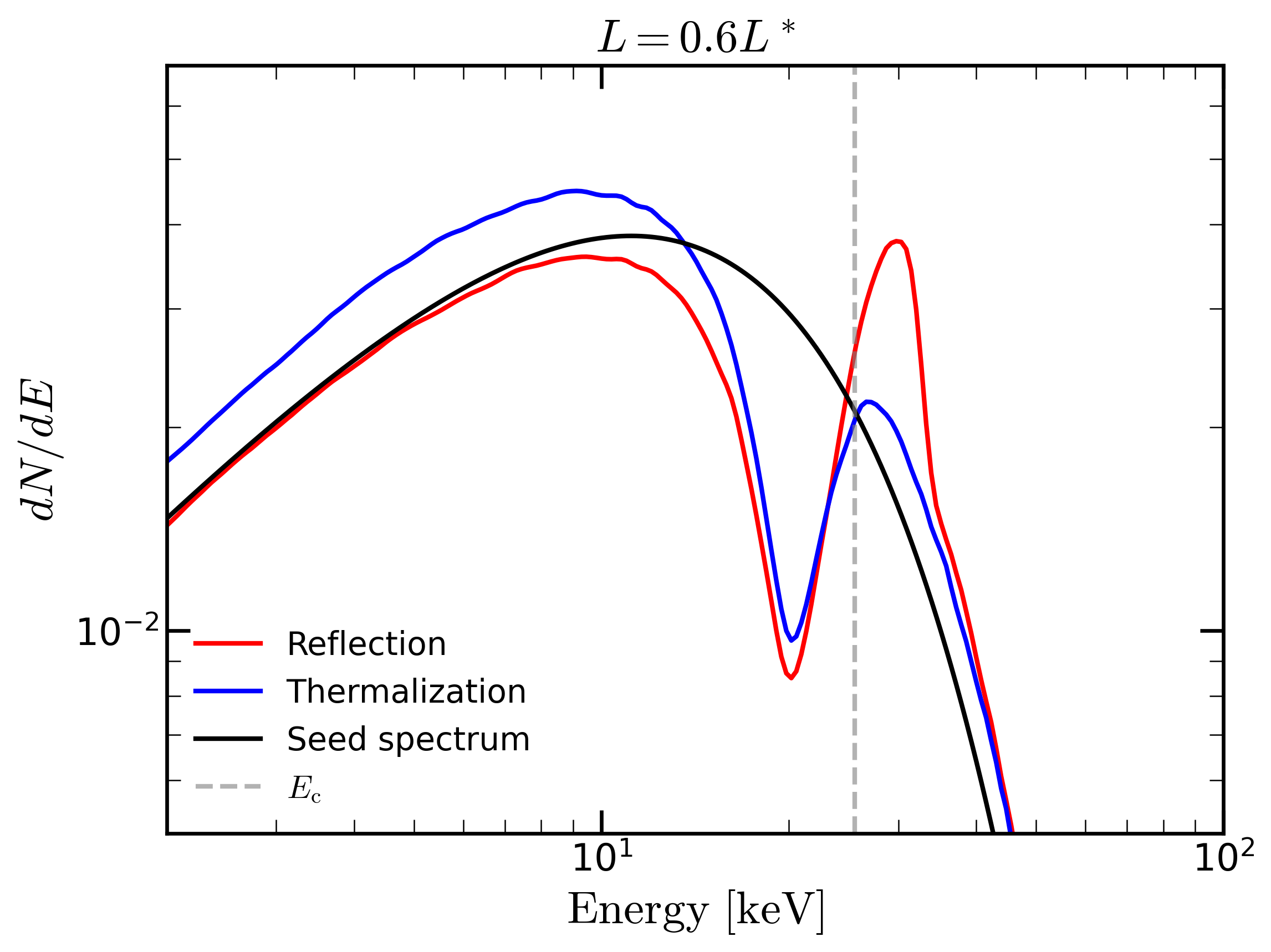}
\caption{Comparison of angle-averaged spectra with different boundary conditions for a typical source with luminosity $L=0.6L^*$ and $E_{\mathrm{c}}=25.55~\mathrm{keV}$ (vertical gray dashed line) . 
The solid black line represents the seed blackbody spectrum with $k_{\mathrm{B}} T=5.9~\mathrm{keV}$. 
The solid red (blue) line shows the emergent spectrum for the Reflection (Thermalization) case. 
The line wings exhibit differences; the red (blue) wing dominates in the Thermalization (Reflection) scheme. 
The centroid energy of the CRSF appears to be independent of the boundary condition. }
\label{fig:spectrum_new}
\end{figure}

In this appendix, we investigate the effect of the boundary condition at the base of the accretion column on the CRSF properties. To do so, we repeated the calculations presented in Sect.~\ref{sec:results}, but this time we used the Thermalization scheme instead of the Reflection one. In other words, photons hitting the base of the accretion column are not reflected back; rather, they get absorbed and are re-emitted as seed photons, following the initial blackbody distribution.
This change is expected to modify the shape of the wings of the line, but not its position, as the scattering mechanism remains the same. In particular, an enhancement of the red wing is expected, as each time a photon packet reaches the base its energy is resampled and $k_{\mathrm{B}}T < E_{\mathrm{c}}$. This process effectively removes photons from the blue wing and redistributes them to the left of the absorption line. 

In Fig.~\ref{fig:spectrum_new}, we compare the angle-averaged spectra obtained using the two different boundary conditions for luminosity $L=0.6L^*$. It is evident that when the Thermalization scheme is implemented (blue line), the red wing dominates over the blue one, whereas for the Reflection scheme (red line), the opposite is observed. However, the position of the absorption line is the same in both cases; the centroid energy of the CRSF is insensitive to the boundary condition. 

\begin{figure}[t]
\centering
\includegraphics[width=0.95\hsize]{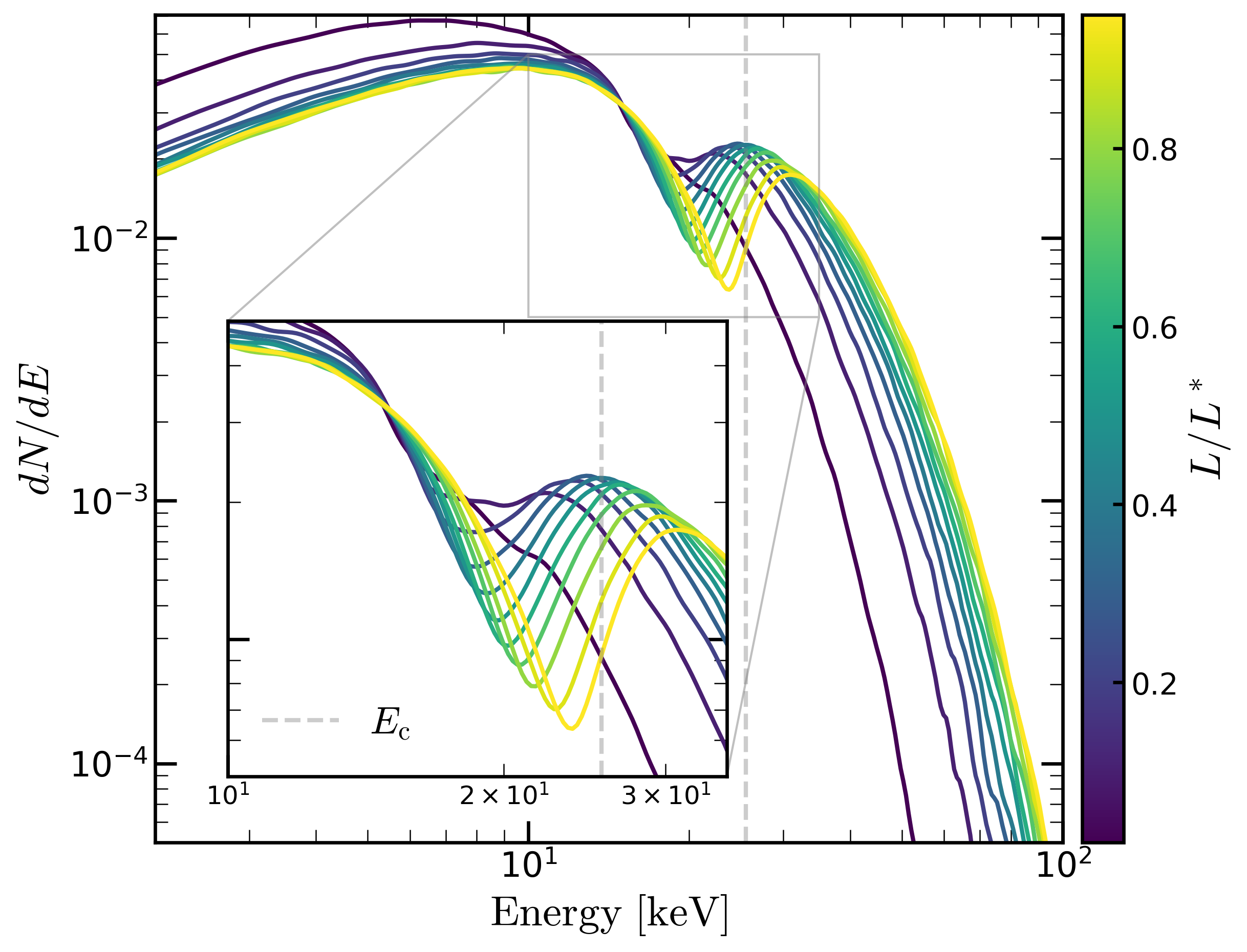}
\caption{Same as Fig.~\ref{fig2}, but for the Thermalization scheme. The luminosity dependence of the CRSF properties is the same.}
 \label{fig99}
\end{figure}

In analogy with Fig.~\ref{fig2}, Fig.~\ref{fig99} shows the resulting spectra for 11 different luminosities $L$. The positive correlation between the position of the CRSF and luminosity is again evident. The Thermalization scheme suppresses the blue wing and enhances the red wing, as described qualitatively above. Even at high luminosities, the blue wing remains relatively weak.

\begin{figure*}[t] 
    \centering
    \begin{subfigure}[b]{0.49\textwidth}
        \centering
    \includegraphics[width=0.95\linewidth]{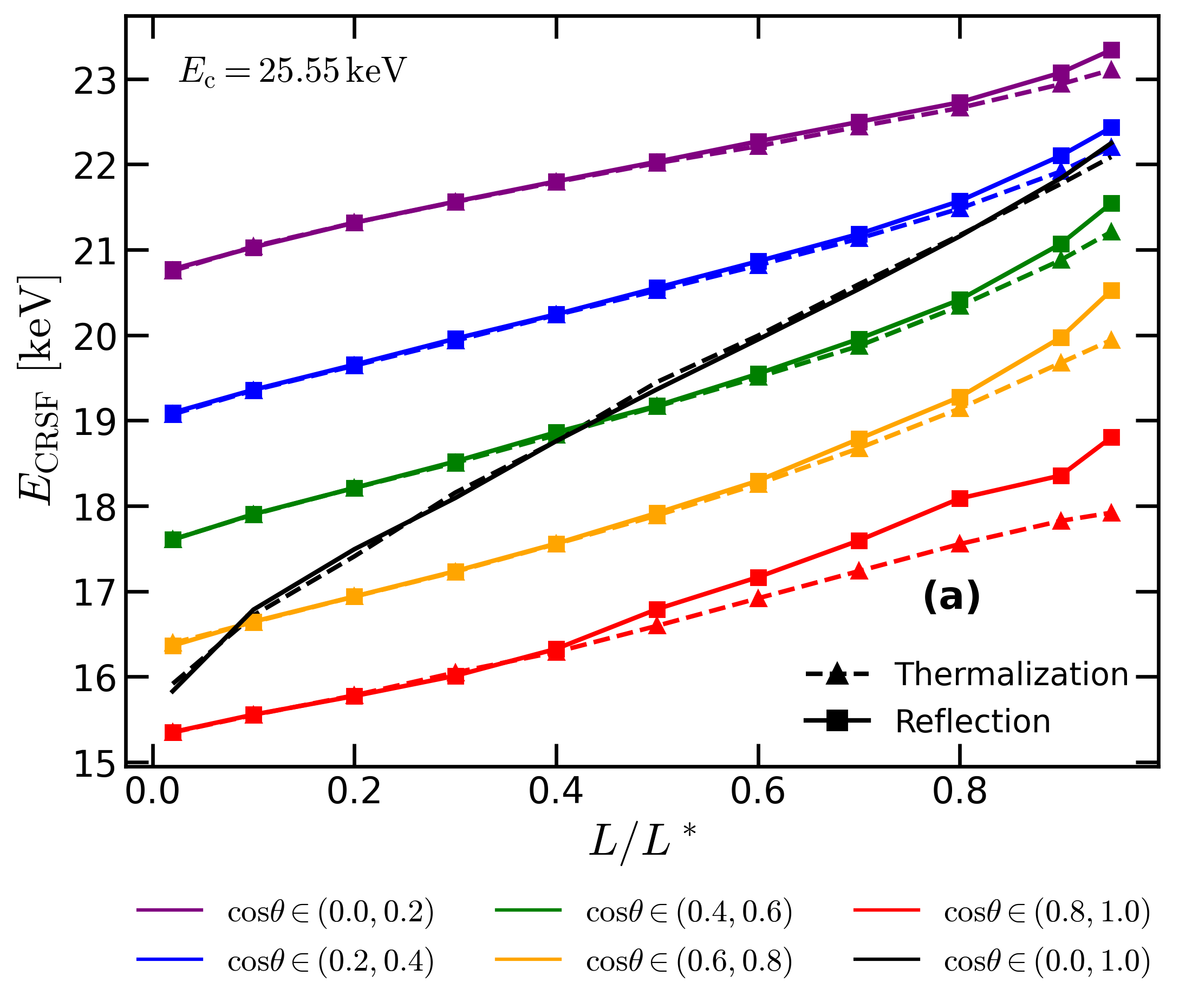}%
    \phantomcaption
        \label{fig:E_comparison}
    \end{subfigure}
    \hfill 
    \begin{subfigure}[b]{0.49\textwidth}
        \centering
    \includegraphics[width=0.95\linewidth]{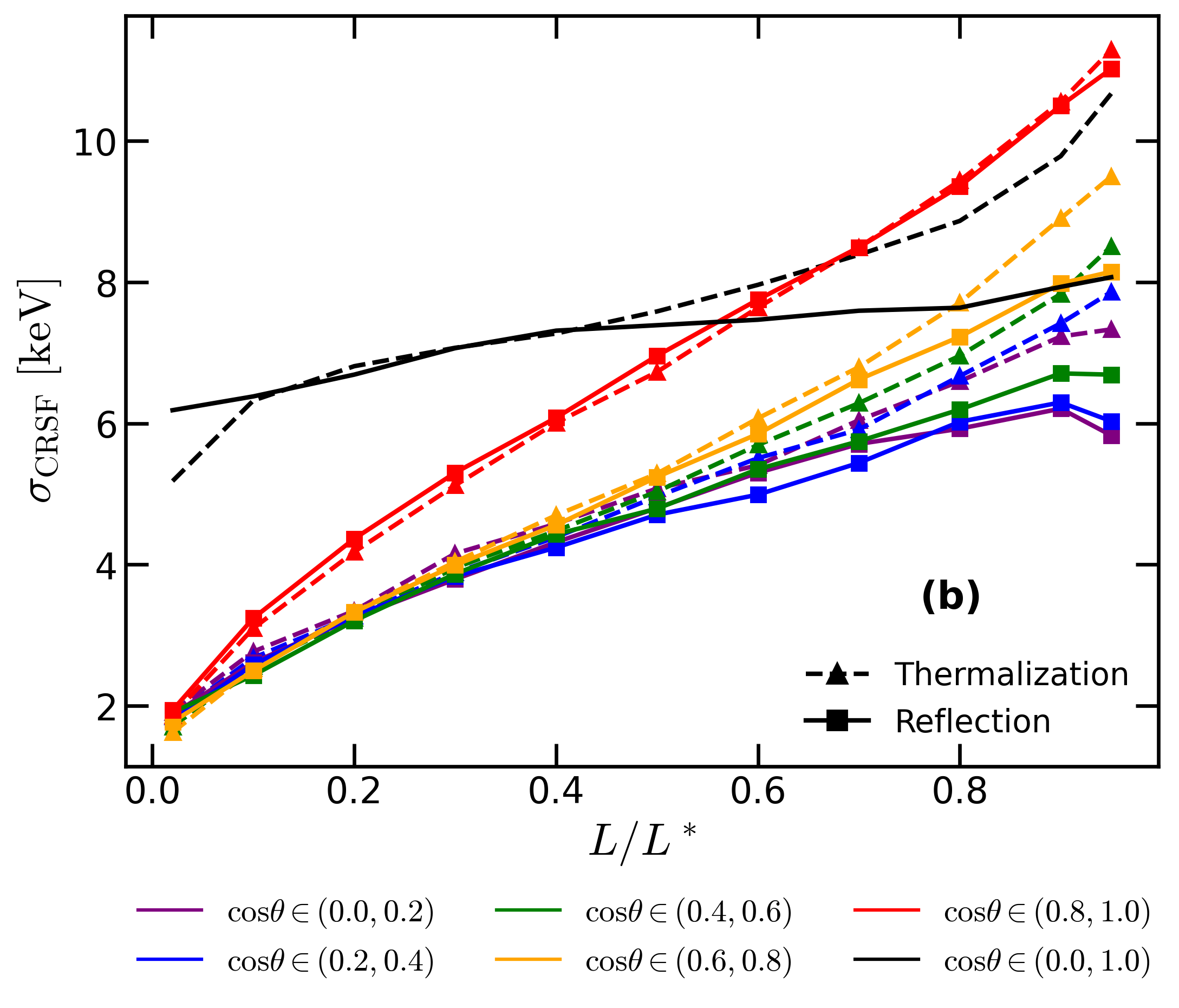}%
    \phantomcaption
        \label{fig:width_comparison}
    \end{subfigure}
    \caption{Same as Fig.~\ref{fig:combined_horizontal}, but including the results obtained using the Thermalization scheme in triangles connected with dashed lines. While the energy of the cyclotron feature is independent of the boundary scheme, the line width shows deviations at high luminosities. This is expected, as the redistribution of photons into the line wings alters the CRSF shape, affecting its width.}
    \label{fig:appendix-comparison}
\end{figure*}

Figure~\ref{fig:appendix-comparison} is the same as Fig.~\ref{fig:combined_horizontal}, but also includes the results obtained using the Thermalization scheme in triangles connected with dashed lines. From  Fig.~\ref{fig:E_comparison}, it is evident that the centroid energy is not affected by the scheme used, regardless of the luminosity and/or viewing angle. 

In contrast, the line width is sensitive to the overall shape of the absorption feature, including its wings; consequently, differences between the two schemes are expected. This is apparent in Fig.~\ref{fig:width_comparison}, where the results for the two schemes diverge at luminosities $L \gtrsim 0.5L^*$. The discrepancy between the dashed (Thermalization) and solid (Reflection) lines increases with luminosity and is most pronounced at small values of $\cos \theta$. The primary cause of this divergence is the effective narrowing of the CRSF in the Reflection scheme due to its prominent blue wing, which is absent in the Thermalization case (see Figs.~\ref{fig:spectrum_new} and \ref{fig99}).
We note, however, that the quantitative determination of the line width is subject to systematic uncertainties. The derived values depend heavily on the modeling of the underlying continuum and the choice of line profile (e.g., Gaussian or Lorentzian).

\section{Absorption feature at low luminosity} \label{App3}
\begin{figure}
\centering
\includegraphics[width=0.95\hsize]{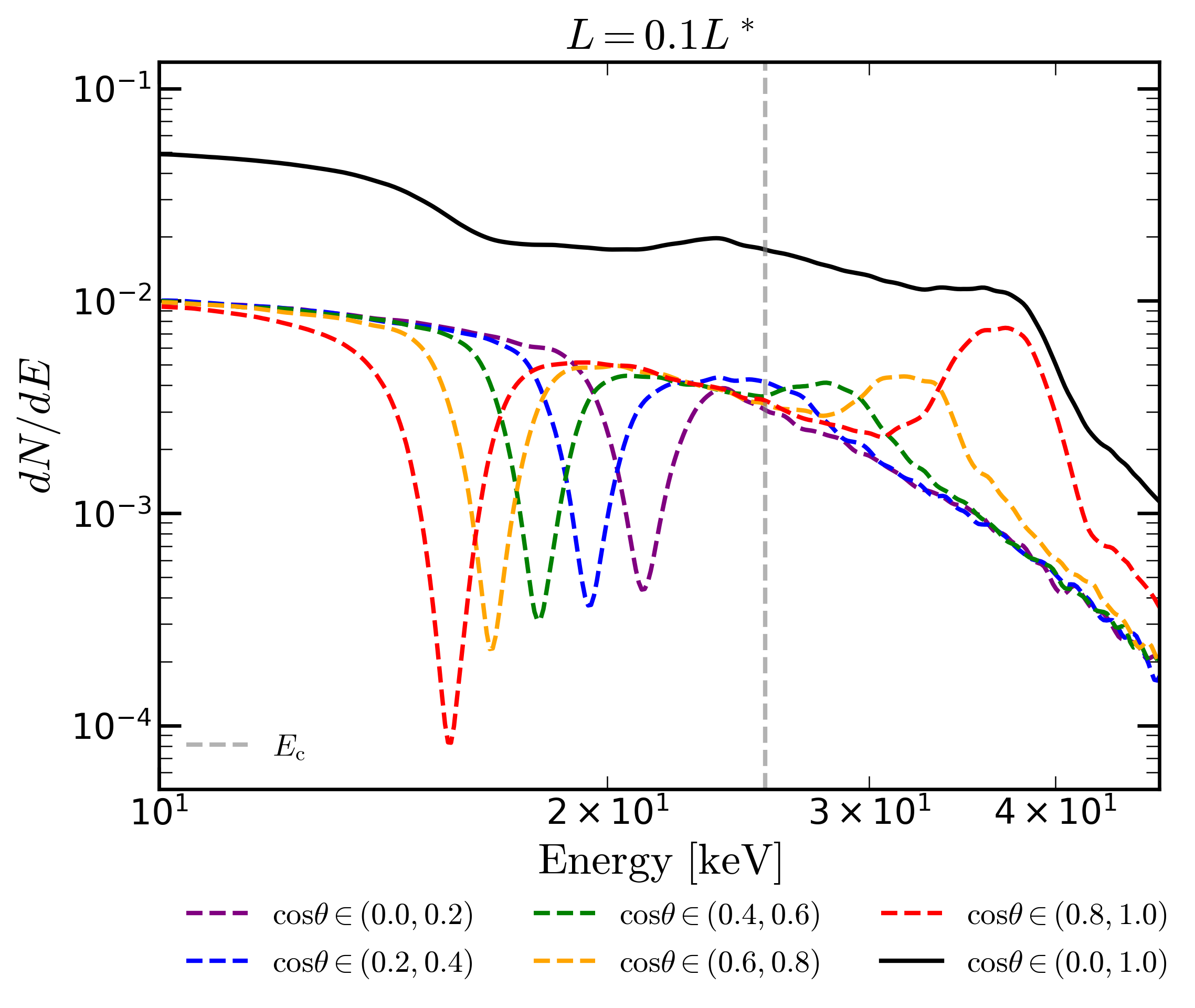}
\caption{Same as Fig.~\ref{fig:angle_resolved}, but for luminosity $L=0.1L^*$. }
\label{fig:angle_resolved_overlap}
\end{figure}

This appendix aims at explaining the lack of a prominent absorption feature in the spectrum at low luminosities in our simulations.
To this end, in Fig.~\ref{fig:angle_resolved_overlap} we present the angle-resolved spectrum for a low luminosity value, $L=0.1 L^*$, calculated using the Reflection scheme. All features seen in Fig.~\ref{fig:angle_resolved} are present here as well: the centroid energy is anti-correlated with $\cos\theta$, and the blue wing becomes increasingly prominent and shifts to higher energies as $\cos\theta$ increases.
    However, in contrast to the $L=0.6 L^*$ case, the superposition of lines emerging at different viewing angles in the low luminosity regime results in a shallow, almost flat CRSF, making it a challenging task to accurately identify the angle-averaged CRSF centroid. Nevertheless, the feature is readily distinguishable after subtraction of the underlying continuum, enabling a reliable determination of the centroid energy.

\section{Calculation of line parameters} \label{App2}

This appendix outlines the methodology used to measure the relevant CRSF parameters: the centroid energy ($E_{\mathrm{CRSF}}$) and the line width ($\sigma_{\rm CRSF}$). The CRSF shape is rather complicated, asymmetric, and deviates significantly from a Gaussian or a Lorentzian profile; thus, we adopted an approach that allows for a robust metric without any bias introduced by fitting specific analytical profiles. We defined the centroid energy as the flux-weighted average energy in the vicinity of the local minimum, and the line width as the full width at half minimum (FWHM) of the absorption feature's depth. 

\begin{figure}
\centering
\includegraphics[width=\hsize]{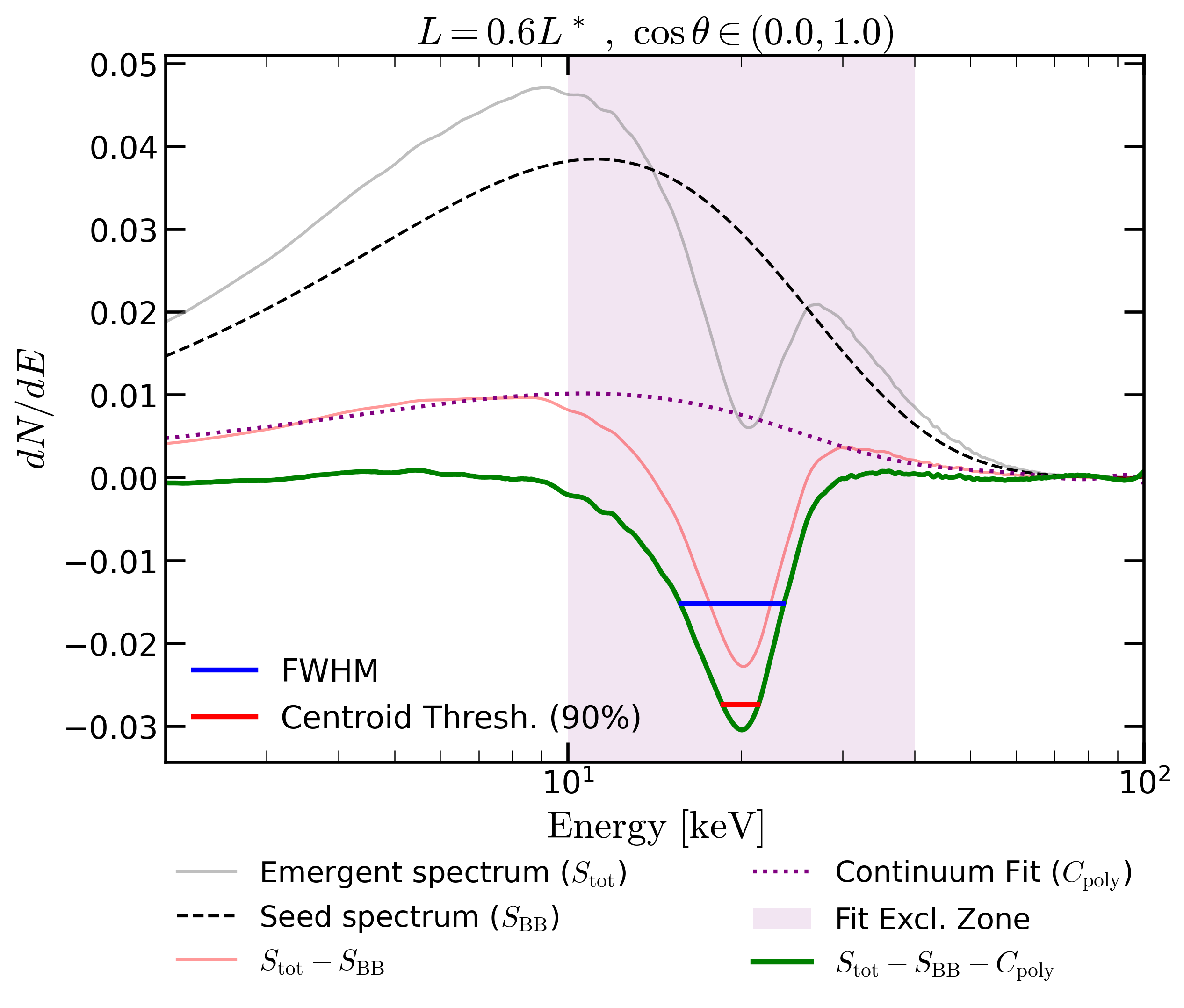}
\caption{Illustration of the procedure used to measure the line parameters for a run with $L=0.6 L^*$. $S_{\mathrm{tot}}$ is the total spectrum in gray, $S_{\mathrm{BB}}$ is the seed blackbody continuum in dashed black, $S_{\mathrm{tot}} - S_{\mathrm{BB}}$ is the residual spectrum in red, and $C_{\mathrm{poly}}$ is the sixth-order polynomial fit of the residual continuum spectrum, excluding the interval around the CRSF (pink shaded region). We measured the line width from the isolated absorption feature ($S_{\mathrm{tot}} - S_{\mathrm{BB}} - C_{\mathrm{poly}}$); green curve) as the FWHM (blue segment). Likewise, we estimated the centroid energy as the flux-weighted average energy within the region around the minimum where the absorption depth is larger than 90\% of the maximum depth (red segment).}
\label{fig:FWHM_demo}
\end{figure}

Figure~\ref{fig:FWHM_demo} illustrates our approach, showing its implementation in a representative angle-averaged spectrum ($\cos\theta\in(0,1)$) for a source with luminosity $L=0.6L^*$, computed using the Thermalization scheme. To estimate the line parameters, we first subtracted the background blackbody continuum ($S_{\rm BB}$; black dashed line) from the total emergent spectrum ($S_{\rm tot}$; solid gray line). Since the resulting residual spectrum ($S_{\rm tot}-S_{\rm BB}$), shown in solid red, does not saturate to zero away from the CRSF, we further fitted the remaining continuum with a sixth-order polynomial fit ($C_{\rm poly}$; dotted purple curve), excluding the energy range (pink shaded rectangular), which is dominated by the absorption feature (i.e., $10\,\mathrm{keV}-40\,\mathrm{keV}$). We, then, subtracted it from the residual spectrum to obtain the isolated CRSF ($S_{\rm tot}-S_{\rm BB}-C_{\rm poly}$; solid green curve). Finally, we measured the line width by computing the FWHM (blue line) and the centroid energy by calculating the flux-weighted average energy within the energy interval around the minimum where the absorption depth exceeds 90\% of the maximum depth (indicated by the red segment in Fig.~\ref{fig:FWHM_demo}). 

We note that the choice of the excluded energy range in the polynomial fit can influence the final measurement; therefore, care must be taken in modeling the continuum to ensure stability. We varied the upper and lower limits of the exclusion zone by $10\%$ to test the sensitivity of our inferences; the derived width and centroid energy change by less than $3\%$. We therefore consider the measurements robust, although a more sophisticated approach would further strengthen their validity.

\end{appendix}

\end{document}